\documentclass[runningheads]{llncs}
\usepackage{graphicx}
\usepackage[dvipsnames]{xcolor}
\usepackage{graphicx}

\usepackage{verbatim}

\usepackage{marvosym}

\usepackage{pifont}

\usepackage{multirow}
\usepackage{booktabs}
\newcommand{\tabincell}[2]{\begin{tabular}{@{}#1@{}}#2\end{tabular}}

\newcommand{\mc}[1]{\mathcal{#1}}

\newcommand{\cu}[1]{\widetilde{#1}}

\begin{document}
%
%\title{Compositional Model Checking of TLA+ Specifications via Interaction-Preserving Abstraction
\title{Compositional Model Checking of Consensus Protocols Specified in TLA+ via Interaction-Preserving Abstraction
    \thanks{Supported by Grant Name (No. xxx). Yu Huang is the corresponding author.}
    %\thanks{Supported by Alibaba Innovative Research (No. 12803031). Hengfeng Wei and Yu Huang are the corresponding authors.}
}
%
%\titlerunning{Abbreviated paper title}
\titlerunning{Compositional Model Checking of Consensus Protocols}
% If the paper title is too long for the running head, you can set
% an abbreviated paper title here
%

\author{Xiaosong Gu\inst{1} \and
    Wei Cao\inst{2} \and
    Yicong Zhu\inst{2} \and
    Xuan Song\inst{2} \and
    Yu Huang\inst{1} \and
   	Xiaoxing Ma\inst{1}}

%\author{Xiaosong Gu\inst{1} \and
%    Second Author\inst{2} \and
%    Third Author\inst{2} \and
%    Hengfeng Wei\inst{1} \and
%    Yu Huang\inst{1}}

%
\authorrunning{X. Gu et al.}

% First names are abbreviated in the running head.
% If there are more than two authors, 'et al.' is used.
%

\institute{
    State Key Laboratory for Novel Software Technology, Nanjing 210023, China \\
    %\email{name@address} 
    \and
    Alibaba Group, Hangzhou 311121, China \\
    \email{mg1933018@smail.nju.edu.cn, \{yuhuang,xxm\}@nju.edu.cn, \\ \{mingsong.cw, zyc141920, qinhuan.sx\}@alibaba-inc.com}
}

%\institute{
%    State Key Laboratory for Novel Software Technology, Nanjing 210023, China \\
%    %\email{lncs@springer.com} 
%    \and
%    Alibaba Group, Hangzhou 311121, China \\
%    \email{mg1933018@smail.nju.edu.cn, yuhuang@nju.edu.cn, \\ \{zyc141920, Author XXX\}@alibaba-inc.com}
%}

%\institute{Princeton University, Princeton NJ 08544, USA \and
%Springer Heidelberg, Tiergartenstr. 17, 69121 Heidelberg, Germany
%\email{lncs@springer.com}\\
%\url{http://www.springer.com/gp/computer-science/lncs} \and
%ABC Institute, Rupert-Karls-University Heidelberg, Heidelberg, Germany\\
%\email{\{abc,lncs\}@uni-heidelberg.de}}

%
\maketitle              % typeset the header of the contribution
%

%--
%\begin{CJK*}{UTF8}{gbsn}

%--
\begin{abstract}

Consensus protocols are widely used in building reliable distributed software systems and its correctness is of vital importance. 
TLA+ is  a lightweight formal specification language which enables precise specification of system design and exhaustive checking of the design without any human effort. 
The features of TLA+ make it widely used in the specification and model checking of consensus protocols, both in academia and industry.
However, the application of TLA+ is limited by the state explosion problem in model checking. Though compositional model checking is essential to tame the state explosion problem, existing compositional checking techniques do not sufficiently consider the characteristics of TLA+.

In this work, we propose the Interaction-Preserving Abstraction (IPA) framework, which leverages the features of TLA+ and enables practical and efficient compositional model checking of consensus protocols specified in TLA+.
In the IPA framework, system specification is partitioned into multiple modules, and each module is divided to the internal part and the interaction part.
The basic idea of the interaction-preserving abstraction is to omit the internal part of each module, such that another module cannot distinguish whether it is interacting with the original module or the coarsened abstract one.

%Then we identity system variables in the specification which convey the interaction among modules. Then the interaction logic is preserved while the logic internal to the modules are omitted. This abstraction process enable efficient compositional checking of each module.

We use the IPA framework to the compositional checking of the TLA+ specification of two consensus protocols Raft and ParallelRaft.
Raft is a consensus protocol which is originally developed in the academia and then widely used in industry.
ParallelRaft is the replication protocol in PolarFS, the distributed file system for the commercial database Alibaba PoloarDB. 
We demonstrate that the IPA framework is easy to use in realistic scenarios and at the same time significantly reduces the model checking cost.

\end{abstract}
%

%--
%\tableofcontents

%--

\section{Introduction}

Consensus algorithms allow a collection of machines to work as a consistent group that can survive partial failures of its members \cite{Lamport01,Ongaro14,Junqueira11}. They play a key role in building reliable large-scale distributed software systems. 
For example, consensus algorithms are used to build coordination services, e.g., Zookeeper \cite{Hunt10} and etcd \cite{Etcd}. Consensus protocols are also used to achieve fault-tolerance for replicated databases, e.g., Chubby \cite{Burrows06,Chandra07}, Spanner \cite{Corbett12}, CosmosDB \cite{Paz18}, and PolarDB \cite{Cao18}.

Since consensus protocols lie in the core of various mission-critical systems, its correctness is of vital importance. 
Traditional software validation techniques are intensively used to improve the reliability of mission-critical systems, e.g. intensive design reviews, code reviews, static code analysis, stress testing, and fault-injection testing \cite{Newcombe15}.
However, deep and subtle bugs are still found to hide in complex concurrent fault-tolerant systems, and are manifested only in rare and extreme cases \cite{Leesatapornwongsa14}. 
It is widely accepted that human intuition is poor at estimating the true probability of supposedly extremely rare combinations of events in systems operating at a scale of millions of requests per second \cite{Newcombe15}.

%\mybreak

%\hy{TLA+被实用}

TLA+ (Tempoal Logic of Actions) is a lightweight formal specification language, especially suitable for design of distributed and concurrent systems \cite{TLA}.
Leveraging simple math, TLA+ can express concepts much more elegantly and accurately than a programming language can.
%What makes TLA+ more suitable for this than, say, Python? Python is designed to be run, and it is limited to what a computer can do. TLA+, though, is designed to be explored. By leveraging simple math, it can express concepts much more elegantly and accurately than a programming language can. 
Specifying a system in TLA+ forces you to be precise in what you actually want.
By unambiguously writing your specification, you understand it better. Problems become obvious even without further exploration. %the model checker. 
More importantly, unlike programming languages, e.g. Java and Go, which are designed to be run and are limited to what a computer can do, TLA+ is designed to be explored.
% instead of being compiled or interpreted, TLA+ is designed to be explored automatically,  without additional human efforts. 
We use a model checker TLC to execute every possible behavior of our specification without additional human efforts. 

The features discussed above make TLA+ widely used in both academia and industry. For example, Paxos and Raft are formally specified and checked using TLA+ \cite{Paxos-TLA,Raft-TLA}, and TLA+ specifications for Zookeeper is under development \cite{ZK-TLA}.
TLA+ is extensively used by Amazon Web Services to help solve deeply-hidden design problems in critical systems \cite{Newcombe15}. 
PolarFS is using TLA+ to precisely document the design of its ParallelRaft protocol, in order to effectively guarantee the reliability and maintainability of the protocol design and implementation \cite{Gu21}.
MongoDB further leverages the formally specified design, verified by model checking, to conduct model-based test case generation and model-based trace checking on large scale system implementations \cite{Davis20}.
% specification and model checking of TLA+ specifications to use the ``extreme modeling" paradigm, i.e., model and implementation are developed in parallel \cite{Davis20}. 

The programmer can view TLA+ specifications as ``runnable designs", which can be machine checked without additional human effort.
However, the model checking of TLA+ specifications is cursed by the notorious state explosion problem \cite{Clarke00}, which limits the scale of checking and restricts the usefulness of TLA+ specifications.
Putting it in another way, increasing the scale of checking can greatly improve the confidence of the system developers that the system does not have bugs pertaining to the complexities and subtleties of fault-tolerant distributed protocol design.

Compositional model checking is essential to increasing the scale of model checking of large distributed systems. 
It addresses the state explosion problem by verifying the individual components without considering the whole system.
Effectiveness of these methods depends on whether an coarse enough (to reduce the checking cost) yet accurate enough (to ensure the correctness of checking) context can be found for each component such that all the essential behavior of that component can be checked.
However, existing compositional checking techniques do not sufficiently consider the characteristics of TLA+ specifications, and are thus not applicable or efficient in model checking of TLA+ specifications.

In TLA+, we model a distributed system in terms of a single global state. This is a simple but generally useful way to model distributed algorithms and systems, as backed by the wide use of TLA+ in both academia, open-source communities and industry.
This salient feature of TLA+ specifications can be utilized to enable efficient compositional model checking.
Moreover, TLA+ is a lightweight formal method. After the specification is given, its model checking is fully automatic. The compositional verification should also be automatic. Formal reasoning after the model checking of each component is not acceptable for the intended users of TLA+. 

Toward the challenges above, we propose the Interaction-Preserving Abstraction (IPA) framework, which is aimed at practical and efficient compositional model checking of TLA+ specifications of realistic distributed consensus protocols.
The framework addresses the challenges above in three steps:
%--
\begin{enumerate}
    \item We divide the system specification in TLA+ into function modules. 
    Each module consists of some actions implementing a specific function. 
    The division is mainly derived from the natural modularity in system design and implementation, which usually has high cohesion and low coupling.
    More importantly, toward the objective of efficient compositional model checking, each module can be divided into two parts: the \textit{internal part} within the scope of one module and the \textit{interaction part} handling interaction with other modules. 
    
    \item We abstract away all the internal logic of each module, while only preserve the interaction logic.
    %The basic rationale of the IPA framework is to abstract away all the internal logic of each module, while only preserve the interaction logic.
    To model check each module separately, we use the abstracted specification of all other modules as the execution context of the module being checked.
    %The key in division is to differentiate system variables which are involved in the interaction from this are only involved in action within a module.
    We provide constraints on the abstraction to ensure that the abstraction preserves the interaction. The constraints are straightforward to check for the specification developers.
    
    \item We provide correctness proof of the compositional checking based on our IPA framework.
\end{enumerate}

We apply the IPA framework to reduce the model checking cost for the specifications of two consensus protocols: Raft and ParallelRaft (PRaft in short). 
Raft is a consensus protocol which is originally developed in the academia and then widely used in industry.
PRaft is the replication protocol in PolarFS, the distributed file system for the commercial database Alibaba PoloarDB \cite{Cao18}. The design of PRaft is derived from Raft and Multi-Paxos \cite{Gu21}.
The case study shows that there are intuitive patterns to conduct the interaction-preserving abstraction, utilizing the characteristics of consensus protocols.
The case study also shows that it is intuitive to guarantee interaction-preservation of the abstraction. Moreover, the constraints in the IPA framework can be conveniently employed to double check the interaction-preservation.
Experimental evaluation shows that the cost for direct checking is up to about 300 times of the cost for compositional checking using our IPA framework.
%10\% $\sim$ 90\% checking cost of direct checking can be saved using compositional checking via our IPA framework.

The rest of this work is organized as follows. 
Section \ref{Sec: Fmk} overviews the IPA framework and Section \ref{Sec: Chk} presents the formal definition.
Section \ref{Sec: Case} presents the case study.
Section \ref{Sec: RW} reviews the related work. 
In Section \ref{Sec: Concl}, we summary this work and discuss the future work.

%--
%--
%\section{Interaction-Preserving Abstraction Framework} \label{Sec: Fmk}
\section{IPA Framework Overview} \label{Sec: Fmk}

The Interaction-Preserving Abstraction (IPA) Framework is designed to enable efficient compositional model checking of TLA+ specifications. 
In this section, we first introduce the characteristics of TLA+ specifications. 
Then we present the workflow to use the IPA framework.

%--------------------------------------------------------------------------------
\subsection{TLA+ Basics}

In the TLA+ specification language, a system is specified as a state machine by describing the possible initial states and the allowed state transitions called $Next$. 
Specifically, the system specification contains a set of system variables $V$. A \textit{state} is an assignment to the system variables. $Next$ is the disjunction of a set of actions $a_1\lor a_2\lor \cdots \lor a_p$, where an \textit{action} is a conjunction of several clauses $c_1\wedge c_2\wedge \cdots c_q$.  A \textit{clause} is either an \textit{enabling condition}, or a \textit{next-state update}. An enabling condition is a state predicate which describes the constraints the current state must satisfy, while the next-state update describes how variables can change in a step (i.e., successive states).

%a predicate of either current state or both current state and next state. We define \textit{enabling condition} of an action to be the conjunction of all predicate clauses of current state. Predicates that involve next state essentially specifies how system variables are assigned by the action, so we define \textit{state update} to be the conjunction of all such clauses. Thus, an action is composed of enabling condition and state update.

%%%%%
%The value assigned to each variable is calculated from values of variables in current state. Usually an action does not modify all system variables, those that are not modified by the execution of the action are noted as \textsf{UNCHANGED} in the assignment.

Whenever every enabling condition $\phi_a$ of an action $a$ is satisfied in a given ``current" state, the system can transfer to the ``next" state by executing $a$, assigning to each variable the value specified by $a$. 
We use ``$s_1\stackrel{a}{\rightarrow}s_2$" to denote that the system state goes from $s_1$ to $s_2$ by executing action $a$, and $a$ can be omitted if it is obvious from the context.
Such execution keeps going and the sequence of system states forms a trace of system behavior. %All such traces are all possible system behaviors. 

TLA+ has a model checker named TLC which builds a finite state model of TLA+ specifications for checking invariance safety properties (in this work, we do not consider liveness properties). 
TLC first generates a set of initial states satisfying the specification, and then traverses all possible state transitions. 
%Execution stops when all state transitions lead to states which have already been discovered. 
If TLC discovers a state which violates an invariance property, it halts and provides the trace leading to the state of violation.
Otherwise, the system passes the model checking and is verified to satisfy the invariance property.

%%%%%
%TLC provides a method of declaring model symmetries to defend against combinatorial explosion \cite{} [14]. It also parallelizes the state exploration step, and can run in distributed mode to spread the workload across a large number of computers \cite{} [20]. As an alternative to exhaustive breadth-first search, TLC can use depth-first search or generate random behaviors. 

In TLA+, correctness properties and system designs are just steps on a ladder of abstraction, with correctness properties occupying higher levels, systems designs and
algorithms in the middle, and executable code and hardware at the lower levels \cite{Newcombe15}.
%TLA+ is intended to make it as easy as possible to show a system design correctly implements the desired correctness properties, through the TLC model checker.
This ladder of abstraction helps designers manage the complexity of real-world systems. Designers may choose to describe the system at several ``middle" levels of abstraction, with each lower level serving a different purpose (such as to understand the consequences of finer-grain concurrency or more detailed behavior of a communication medium). The designer can then verify that each level is correct with respect to a higher level. The freedom to choose and adjust levels of abstraction makes TLA+ extremely flexible.
%Refinement is a key built-in feature of TLA+. A system can be modeled using TLA+ at different levels of abstraction and refinement is used to prove that a low-level specification correctly implements a high level one, thus ensuring the implementation preserves the desired high-level properties. 
For example, a low-level specification for leader election mechanism of Raft may accurately describe how an eligible server is selected as leader through voting, while a high-level one may directly assign some eligible server to be leader and leave the details of voting unspecified. 
%TLC can check that whether an specification is a refinement of another one automatically given the refinement mapping.

%------------------------------
\subsection{IPA Workflow}

\begin{figure}[htbp]
    \centering
    \includegraphics[width=0.8\linewidth]{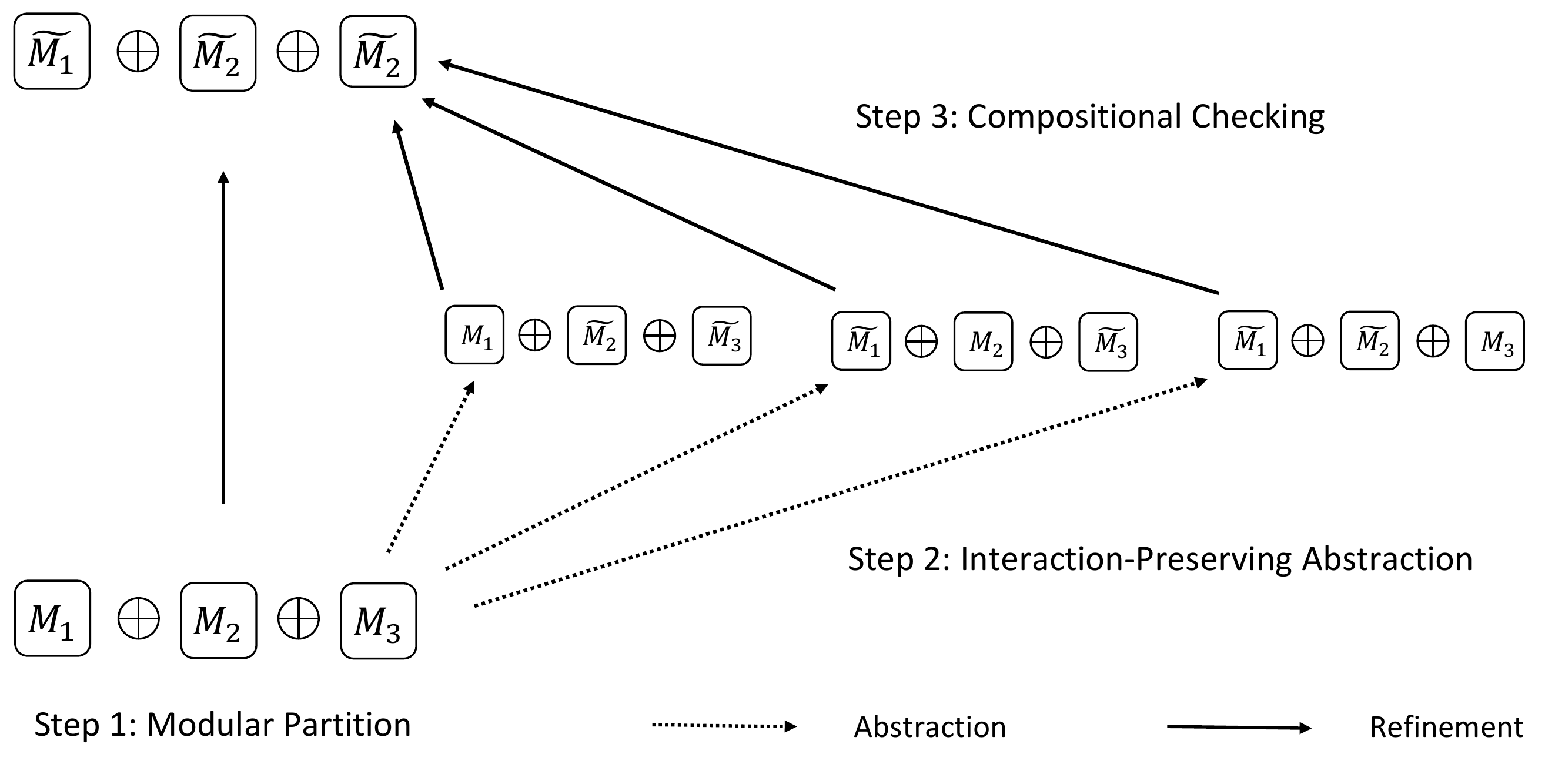}
    \caption{The Interaction-Preserving Framework.}
    \label{F: IPA}
\end{figure}

The basic objective of the IPA framework is to divide the system specification to modules, and model check each module separately, as shown in Fig. \ref{F: IPA}. 
To achieve compositional checking, we abstract away the internal logic of each module and only preserve the interaction logic.
The abstracted modules serve as the execution context for the module to be checked separately.
%--

As shown in Fig. \ref{F: IPA}, we have three levels of specifications. In the lowest level, we have the original specification. In the middle level, we have the compositional specification. In the uppermost level, we have the abstract specification. We assume that the abstract specification has passed the model checking of the correctness property. The main task is to verify that the original specification refines the abstract specification, thus also satisfying the correctness property.
Based on the IPA framework, we only need to model check that the compositional specification refines the abstract specification. 
This will imply that the original specification refines the abstract specification, as we will prove in Section \ref{SubSec: Proof}.
There are basically three steps when applying the IPA framework, as detailed below. 

%----------------------------------------
\subsubsection{Modular Partition of Specification}

Distributed systems are difficult to design and implement.
To control the complexity of system development, it is common practice to design a distributed system as the composition of a collection of function modules.
The modules are expected to have high cohesion and low coupling.
For example, classic consensus protocol Raft can in high-level be divided into two modules: one for leader-based log replication and the other for recovery from failure of the leader. 

%Transaction control algorithm contains at least two modules: atomic commit protocol which ensures atomicity and concurrency control protocol for isolation. 

The IPA framework leverages such modularity.
The system specification is partitioned into multiple modules. Checking each module separately obviously can save the model checking cost significantly. 
However, the critical challenge in this partition process is to handle the unavoidable interaction among modules, and to construct an execution context for each module. 
The context should be sufficiently accurate, i.e., each module should be provided with the ``illusion" that it interacts with other modules, not with the contexts minimized for compositional checking.
The context should also be coarsened enough, otherwise the compositional checking cannot significantly reduce the checking cost.

We enforce the modular partition process by leveraging the characteristics of TLA+ specifications.
In TLA+, we model a distributed system in terms of a single global state. 
This is a generally useful way to model distributed algorithms and systems, as backed by the wide use of TLA+ in the academia, the open-source community and the industry.
As for TLA+ specifications, the key ingredients are the variables and the actions. 
To partition the system specification into modules is just to partition all the actions. Each module is just a subset of actions.
Theoretically, this partition can be arbitrary. 
However, as discussed above, a distributed system is usually based on a modular design. This modularity should be and can naturally be preserved in the TLA+ specification.
Module partition of the TLA+ specification should respect such modularity in system design because the high cohesion and low coupling nature of the design is expected to better reduce the model checking cost.

The interaction among modules is based on read and write of system variables which are shared among modules, similar to global variables shared by multiple functions in C programming.
For example in Raft, the log replication module and the leader election module access common variables (e.g. $term$ and $log$) and have subtle interdependencies, which makes it not feasible to check each module separately.
We identify interaction among modules by identifying \textit{interaction variables}, i.e., system variables which ``convey" the interaction among modules. This enables further interaction-preserving abstraction. Based on the identification of the interaction variables, we can divide the logic of one module into two parts. One is the \textit{internal} part, which just updates information within the module. The other is the \textit{interaction} part, which involves interaction with other modules. 

%----------------------------------------
\subsubsection{Interaction-Preserving Abstraction}

To construct an execution context for each model as required by compositional model checking, we conduct interaction-preserving abstraction for each module. 
As indicated by its name, in this abstraction process, actions which do not interact with other modules are omitted.
The coarsened abstract module is equivalent to the original detailed one, in the sense that other module cannot distinguish the abstract module from the detailed one during interaction.

%The abstraction here is the ``inverse" of refinement. 
%The actions which interact with other modules are preserved. 
%The abstraction essentially simplifies the inner logic of the module while preserving its interactions with other modules. 

When interaction among modules are simple, this abstraction process is often straightforward. For example, a distributed lock service has clear interfaces for other modules no matter how complicated the service is implemented. Therefore, the interaction-preserving abstraction for a lock service is simply the specification of the semantics of the lock service APIs.
However, in many cases, the interaction among modules is much more complex and subtle. For example, the replicated log in Raft is accessed by multiple functionally different modules, e.g. the leader election module and the log replication module. Abstraction of actions manipulating the log is quite non-trivial. The key of the interaction-preserving abstraction process is to identify the system variables which ``convey" the interaction among modules, as detailed in Section \ref{SubSec: Module}.

%--------------------------------------------------------------------------------
\subsubsection{Compositional Checking}

The compositional checking based on the IPA framework is an indirect approach to verifying that the original specification satisfies the correctness property.
Specifically, our main objective is to verify that the original specification refines the abstract specification, assuming that the abstract specification satisfies the correctness property, as shown in the left side of Fig. \ref{F: IPA}.
In our indirect approach, we first conduct the interaction-preserving abstraction for each module and obtain the compositional specification for each module.
Then we model check that each compositional specification refines the abstract specification.
We will prove in Section \ref{SubSec: Proof} that the indirect checking implies the original direct checking.
We will also show in Section \ref{SubSec: Exp} that the indirect checking can significantly save model checking cost.

%Given the interaction-preserving abstraction for each module, we define the \textit{compositional specification} for each module, which is composed of the original specification for this module and the IPA specifications for all other modules. 

%Combining abstractions for all modules, we get an abstracted specification for the whole system. 

%Essentially, we want to check that the original specification refines abstracted specification, so that invariance properties checked for abstracted specification also hold for original specification.
%Here we assume that the top layer of abstract specifications can be model checked and the invariance properties hold.
%Instead of running model checking experiments directly on this refinement which is usually difficult since low level details greatly increase checking cost, we can perform compositional checking, which checks the refinement from a hybrid specification to the abstracted specification. 

%We prove in Section \ref{Sec: Chk} that if all compositional specifications refine the abstract specification, it implies that refinement from the original specification to the abstract specification also holds.

%--
%--
\section{Compositional Model Checking based on Interaction- Preserving Abstraction} \label{Sec: Chk}

In this section, we present formal description of the IPA framework. We first describe how to divide a system specification in TLA+ into modules and how to capture the interaction among modules. Then we describe how to conduct interaction-preserving abstraction for each module. Third, we prove that passing the compositional checking via the IPA framework implies passing the direct checking of the original specification.

%----------------------------------------
\subsection{Modules and Interactions among Modules} \label{SubSec: Module}

A system usually consists of several modules, each implementing some specific function. 
For example, the consensus protocol Raft may be divided into two modules: \textit{log replication}, which describes how the nodes reach consensus as instructed by a leader and \textit{leader election}, which specifies how a new leader is elected when the original leader fails. For the TLA+ specification of a distributed system design, we define: 
%--
\begin{definition}[module]
    A module is a collection of actions. All the modules form a partition of all the actions in the specification.
\end{definition}

\noindent Modules interact with each other through the system variables.
To capture this, we first define the dependency variable of an action and that of a module:
%--
\begin{definition}[dependency variable] \label{Def: DepVar}
    Suppose module $M=\{a_1,a_2,\cdots ,a_m\}$, dependency variables of $M$, denoted as $\mathcal{D}_{M}$, is obtained recursively according to the following rules:
    %--
    \begin{enumerate}
        \item For any action $a_i\in M$, its dependency variables $\mathcal{D}_{a_i}$ are the variables which appear in some enabling condition $\phi_{a_i}$ of $a_i$. % Dependency variables of an action is the set that is related to whether the action is enabled.
        \item $\bigcup\limits_{1\leq i\leq m}\mathcal{D}_{a_i}\subseteq \mathcal{D}_M$. That is, the dependency variables of each action in $M$ belong to $\mathcal{D}_M$.
        \item For any $v \in \mathcal{D}_{M}$ and any action $a_i\in M$, if the next-state update of $a_i$ assigns to $v$ a value calculated from multiple variables (denoted by variable set $V_{dep}$), then $V_{dep} \subseteq \mc{D}_M$. This is due to the fact that the dependency relation is transitive, i.e., if $M$ depends on some variable $v$ and $v$ depends on another variable $w$, then $M$ also depends on $w$.
    \end{enumerate}
\end{definition}

%%%%%
% \noindent Note that the variables present in the next-state update of $a$ may not belong to $\mathcal{D}_a$. 
%\hy{对别的模块的变量x的写。我不依赖x。x=f(y,z)，我也不依赖y，z。}

\noindent Given the definitions above, we can now say that module $M_i$ interacts with $M_j$ by modifying $\mc{D}_{M_j}$.

%\hy{下面的论述和上面的rule3并不冲突。补一个文字解释？}

The notion of the dependency variable alone is not sufficient to  capture the interaction among modules, since even if $D_{M_i}$ are not modified by some action in $M_j$, $M_i$ may still be affected indirectly.
Suppose $x \in \mathcal{D}_{M_i}$, %is in dependency variables of module $M_i$, 
an action in another module $M_j$ assigns to $x$ the value of $y$ (note that $y$ will not be added to $\mc{D}_{M_i}$ by the Rule 3 in Definition \ref{Def: DepVar}, since $x$ is assigned the value of $y$ in module $M_j$, not in $M_i$). 
%Variable $y$ is not in $\mathcal{D}_{M_i}$. 
In this case, any assignment to $y$ may also change the value of $x$ in subsequent actions. 
To capture such indirect interactions between modules, we define the set of interaction variables $\mc{I}$:
%--
\begin{definition}[interaction variable]
    Suppose the specification contains $k$ modules: $M_1,\cdots,M_k$. %$V$ is the set of system variables. 
    The set of interaction variables $\mathcal{I}$ is calculated recursively according to the follwing rules:
    %--
    \begin{enumerate}
        \item $\bigcup\limits_{1\leq i<j\leq k}(\mathcal{D}_{M_i}\cap \mathcal{D}_{M_j}) \subseteq \mathcal{I}$. That is, variables which are dependency variables of multiple modules belong to $\mathcal{I}$.
        
        \item For any $v \in \mathcal{I}$ and any module $M_i$, if an action $a \in M_i$ assigns to $v$ a value calculated from multiple variables (denoted by set $V_{intr}$), then add all variables in $V_{intr}\setminus \mathcal{D}_{M_i}$ to $\mathcal{I}$. That is, the value assigned to an interaction variable by any action in $M_i$ should be calculated from values of variables in interaction variables or dependency variables of the module, i.e., $\mathcal{I} \cup \mathcal{D}_{M_i}$.
        
        \item For any variable $v \in \mathcal{D}_{M_i}\setminus \mathcal{I}$ in any module $M_i$, if an action assigns to $v$ a value calculated from multiple variables (denoted by set $V'_{intr}$), then add all variables in $V'_{intr}\setminus \mathcal{D}_{M_i}$ to $\mathcal{I}$. That is, the value assigned to a ``internal'' variable of $M_i$ by any action should be calculated from values of interaction variables or from values of dependency variables of the module, i.e., $\mathcal{I}\cup \mathcal{D}_{M_i}$.
    \end{enumerate}
\end{definition}

\noindent Note that in Rule 1 of this definition, we are a bit conservative. Some variable $x$ in both $\mc{D}_{M_i}$ and $\mc{D}_{M_j}$ may not convey any interaction between $M_i$ and $M_j$. However, in practice this case is rare (see details of our case study in Section \ref{Sec: Case} and Appendix \ref{A: Case-Raft} and \ref{A: Case-PRaft}) and we ignore this case to make our definition concise and easy to use.

Given the definition of the interaction variable, it is straightforward to verify that: for any two different modules $M_i$ and $M_j$, $(\mathcal{D}_{M_i} \setminus \mathcal{I}) \cap \mathcal{D}_{M_j} = \emptyset$.
%\gxs{是补一个附录证明，还是也可以省，我直接在正文说：显然可以验证。还是连引理环境都省了，直接在正文里写。}
We define the internal variables of module $M_i$, denoted as $\mathcal{L}_{M_i}$, to be $\mathcal{D}_{M_i}\setminus \mathcal{I}$. Intuitively, if all variables but $\mathcal{L}_{M_i}$ stay unchanged in an action in $M_i$, then this action has no effect on other modules. 

%--------------------------------------------------------------------------------
\subsection{Interaction-Preserving Abstraction for Each Module}

The main objective of our IPA framework is to enable separate model checking of each module, in order to reduce the cost for direct checking of the original specification.
The critical challenge is to construct an execution context for each module, such that all the behaviors in the module can be checked separately.

To this end, we conduct interaction-preserving abstraction for each module. Suppose we have $k$ modules $M_1, M_2, \cdots, M_k$. The abstraction of each module $M_i$ is denoted by $\cu{M_i}$.
When we check module $M_i$ separately, the abstractions of all other modules, i.e. all $\cu{M_j}$ ($j\neq i$), serve as the execution context of $M_i$

The key in the abstraction process is to omits internal details of every module as much as possible, and more importantly, the logic concerning interaction among modules must be preserved.
We need to ensure that one module cannot distinguish whether it is interacting with the original specification of other modules or the abstracted specifications.

%----------------------------------------
\subsubsection{Formal Definition of Interaction-Preservation}

We now present the formal definition of the interaction-preserving abstraction. 
The abstraction process obtaining each $\cu{M_i}$ may introduce new variables and actions. We can define the dependency variables of the abstracted module $\mc{D}_{\cu{M_i}}$ in the same way, according to Definition \ref{Def: DepVar}.
The abstracted specification $\cu{M_i}$ should satisfy the following constraints:
%--
\begin{enumerate}
    \item As $\cu{M_i}$ is the abstraction of $M_i$, the dependency variables of $\cu{M_i}$ should not intersect with the local variables of other modules. Formally, $\mathcal{D}_{\cu{M_i}}\subseteq \mathcal{I}\cup \mathcal{D}_{M_i}$. 
    
    \item For updates of interaction variables in $\mc{I}$, the value assigned to any interaction variable by any action in $\cu{M_i}$ should be calculated from values of interaction variables or those of dependency variables of the module, not from values of internal variables of other modules. That is, for any variable $v\in \mathcal{I}$, the value assigned to $v$ by any action of $\cu{M_i}$ is calculated from values of  $\mathcal{D}_{\cu{M_i}} \cup \mathcal{I}$.
    
    \item For updates of internal variables of each abstracted module, the value assigned to any internal variable of the module by any action should be calculated from values of interaction variables or those of dependency variables of that module, not from values of internal variables of other modules. That is, for any variable $v\in \mathcal{L}_{\cu{M_i}}$, the value assigned to $v$ by any action is calculated from values of $\mathcal{I}\cup \mathcal{D}_{\cu{M_i}}$.
    
    \item Abstraction of any module preserve all actions whose effect can be ``perceived'' by other modules. This requires that there is a mapping $f_i: M_i\rightarrow \cu{M_i}$, such that for any action $a\in M_i $ and any module $M_j(j\neq i)$, $f$ and $f_i(a)$ modify the values of $\mc{D}_{\cu{M_i}} \cup \mathcal{D}_{M_j}\cup \mathcal{I}$ in the same way. Note that if action $a$ only changes the values of $\mathcal{L}_{M_i}$ and leave all other variables unchanged, then $f_i(a)$ may be void. Specially, $f_i(a)$ preserves all assignment clauses to variables in $\mathcal{L}_{M_j}$ syntactically.
\end{enumerate}

\noindent According to the constraints above, some internal variables as well as actions that only modifies these variables are omitted in the abstraction.

%\hy{解释一下，上面的规则，实际使用时不会太麻烦，如案例section所示。上述规则仅仅用于检查。}

%----------------------------------------
\subsubsection{Three Layers of Specifications}

Initially, we are given the original specification which is partitioned into modules: $S = \bigcup\limits_{1\leq i\leq k}M_i$.
In order to define the compositional specification for each module, i.e., original specification for one module and abstracted specification for all other modules, we need to define the variables and actions of the compositional specification. 

Define $C_i$ to be the specification that combines $M_i$ and every $\cu{M_j}(j\neq i)$, i.e. $C_i=(\bigcup\limits_{j\neq i}\cu{M_j})\cup M_i$. % The actions of $C_i$ are actions of $M_i$ and actions of $\cu{M_j}(j\neq i)$ put together, i.e., $C_i=\bigcup\limits_{j\neq i}\cu{M_j}\cup M_i$. 
Let the system variables of specification $C_i$ be $V_{C_i} = \mathcal{I}\cup \mathcal{D}_{M_i} \cup (\bigcup\limits_{j\neq i} \mathcal{D}_{\cu{M_j}})$. 
It is obvious that variables not in $V_{C_i}$ are irrelevant to the execution of $C_i$ because $V_{C_i}$ contains all the dependency variables of modules in $C_i$ and any assignment to variables of $V_{C_i}$ is calculated from variables in $V_{C_i}$. % We therefore remove all variables not in $V_{C_i}$ from $C_i$. 

Define $A$ to be the specification that combines all abstracted specifications for each module, i.e., $A=\bigcup\limits_{1\leq i\leq k}\cu{M_i}$. 
Variables $V_A=\mathcal{I} \cup \bigcup\limits_{1\leq i\leq k} \mathcal{D}_{\cu{M_i}}$ are all variables that are relevant to the execution of $A$. % We remove all variables not in $V_A$ from $A$. 

%----------------------------------------
\subsubsection{Strong Refinement Relation between Specifications}

By defining the compositional specifications and the abstract specification, we can now circumvent the direct checking (that $S$ refines $A$) using the compositional checking (that every $C_i$ refines $A$).
The original definition of the refinement relation between two protocols only requires that there is a mapping between the traces of two protocols. 
Now in order to enable compositional checking in our IPA framework, we strengthen the definition of the refinement relation with additional requirement on the mapping between actions. Similar enhancement of the refinement relation is also used in the existing work \cite{Bornholt21}.
First we present the formal definition of refinement between protocols:
%--
\begin{definition}[refinement]
    A refinement mapping from protocol $B$ to $A$ assigns to each variable $v$ of $A$ an expression $\bar{v}$, where $\bar{v}$ is defined in terms of variables of $B$. A refinement mapping defines for each state $s$ of $B$ a state $s'$ of $A$ in which the value of each variable $v$ is mapped to the value of $\bar{v}$ in state $s$. 
    
    Protocol $B$ refines $A$ if and only if there is a refinement mapping from $B$ to $A$ such that for each valid trace of $B$: $s_1\rightarrow s_2\rightarrow \cdots\rightarrow s_t$, $s_1'\rightarrow s_2'\rightarrow\cdots\rightarrow s_t'$ is a valid trace of $A$.
\end{definition}

The definition of refinement only requires a mapping from the state space of $B$ to that of $A$. 
In order to get an abstract specification, it is common to omit some actions which are about low level details. In this case, there is an obvious correspondence between actions of specifications. That is, some actions are preserved in both the abstract and the detailed specifications, while some actions are directly omitted (mapping to a void action). 
Given a trace of the original detailed specification, we can use such correspondence between actions to construct a corresponding trace of the compositional specification, and then construct a corresponding trace of the abstract specification. This helps us prove the refinement from the original specification to the abstract specification. 
The detailed proof will be provided in Section \ref{SubSec: Proof}. Now we first define the strong refinement relation to capture the correspondence between actions:
%--
\begin{definition}[strong refinement]
    $B$ strongly refines $A$, denoted by $B\Rightarrow A$, if and only if $B$ refines $A$ and there is a mapping $f(\cdot)$ from actions of $B$ to those of $A$, such that for any valid trace of $B: s_1\stackrel{a_1}{\longrightarrow}s_2\stackrel{a_2}{\longrightarrow}\cdots\stackrel{a_{t-1}}{\longrightarrow}s_t$, $s_1'\stackrel{f(a_1)}{\longrightarrow}s_2'\stackrel{f(a_2)}{\longrightarrow}\cdots\stackrel{f(a_{t-1})}{\longrightarrow}s_t'$ is a valid trace of $A$.
\end{definition}

%\gxs{- 解释之前，首先要介绍整体设定：有三层，如图所示，$S$，$C_i$，$A$。\\
%    - 每个$M_i$到$\cu{M_i}$，记为f，是强精化映射，是人肉搞定的。\\
%    - 对于$S$到$C_i$, $M_i$不用变，是恒等映射，其它的$M_j$是粗化那个映射\\
%    - 对于$C_i$到$A$，$M_i$变成f映射，其它变成恒等映射？}

\noindent It is obvious to see that strong refinement is transitive.

%\hy{名字相同的变量对应，使用identity mapping不知道是否恰当}
%\hy{这里的$V$指原规约中的所有变量，似乎之前没有说明。。}

%We now have three layers of specifications: the original specification $S$, compositional specifications $C_i$ and abstracted specification $A$. All specifications are composed of same modules with different abstraction. 

Given the action mapping $f_i:M_i \rightarrow \cu{M_i}$ for each module $M_i$, we now establish strong refinement relations from $S$ to $C_i(1\leq i \leq k)$, $C_i$ to $A$ and $S$ to $A$, as shown in Fig. \ref{F: IPA}. 
The strong refinement mapping has two parts: the mapping between variables, and the mapping between actions.
According to the definitions of $S, C_i$ and $A$, $V_A\subseteq V_{C_i} \subseteq V$ (here we assume that both variables in the original specification $S$ and variables introduced in the abstract specifications $\cu{M_i}$ for $1\leq i\leq k$ are in $V$), so refinement mappings concerning the variables are identity mappings.

Action mapping from $S$ to $C_i$, denoted as $g_i$, is defined as follows. For any action $a\in S$, if $a\in M_i$, then $g_i(a)=a$. If $a\in M_j(j\neq i)$, then $g_i(a)=f_j(a)$. 
Action mapping from $C_i$ to $A$, denoted as $\bar{g}_i$, is defined in a similar way: if $a \in M_i$, then $\bar{g}_i(a)=f_i(a)$ and if $a\in \cu{M_j}(j\neq i)$, then $\bar{g}_i(a)=a$. 
Action mapping from $S$ to $A$, denoted as $g$, maps each action to its abstracted version: for any action $a\in M_i$, $g(a)=f_i(a)$. It is straightforward to see that for any $i$, $g$ is the composite function of $g_i$ and $\bar{g}_i$, i.e., $\bar{g_i}(g_i)=g$.

%--------------------------------------------------------------------------------
\subsection{Correctness of Compositional Checking} \label{SubSec: Proof}

We have presented the basic workflow using the IPA framework. The basic rationale behind the IPA framework is to use the compositional checking of each $C_i$ to circumvent the direct checking of the original specification $S$. This circumvention is backed by the following theorem:
%--
\begin{theorem}[Correctness of compositional checking] \label{Th: Chk}
    % Suppose specification $S$ is composed of $k$ modules :$M_1,\cdots,M_k$. Each module $M_i$ is abstracted as $\cu{M_i}$. Let $C_i$ be the compositional specification composed of $M_i$ and $\cu{M_j}(j\neq i)$. Let $A$ be the specification with all modules abstracted. 
    % If $\forall i, C_i\Rightarrow A(1\leq i\leq k)$, %under the refinement mappings and action mappings defined above, 
    % then $S\Rightarrow A$.

    $\forall \ 1\leq i \leq k : C_i\Rightarrow A$ implies that $S\Rightarrow A$.
\end{theorem}

\begin{figure}[htbp]
    \centering
    \includegraphics[width=0.8\linewidth]{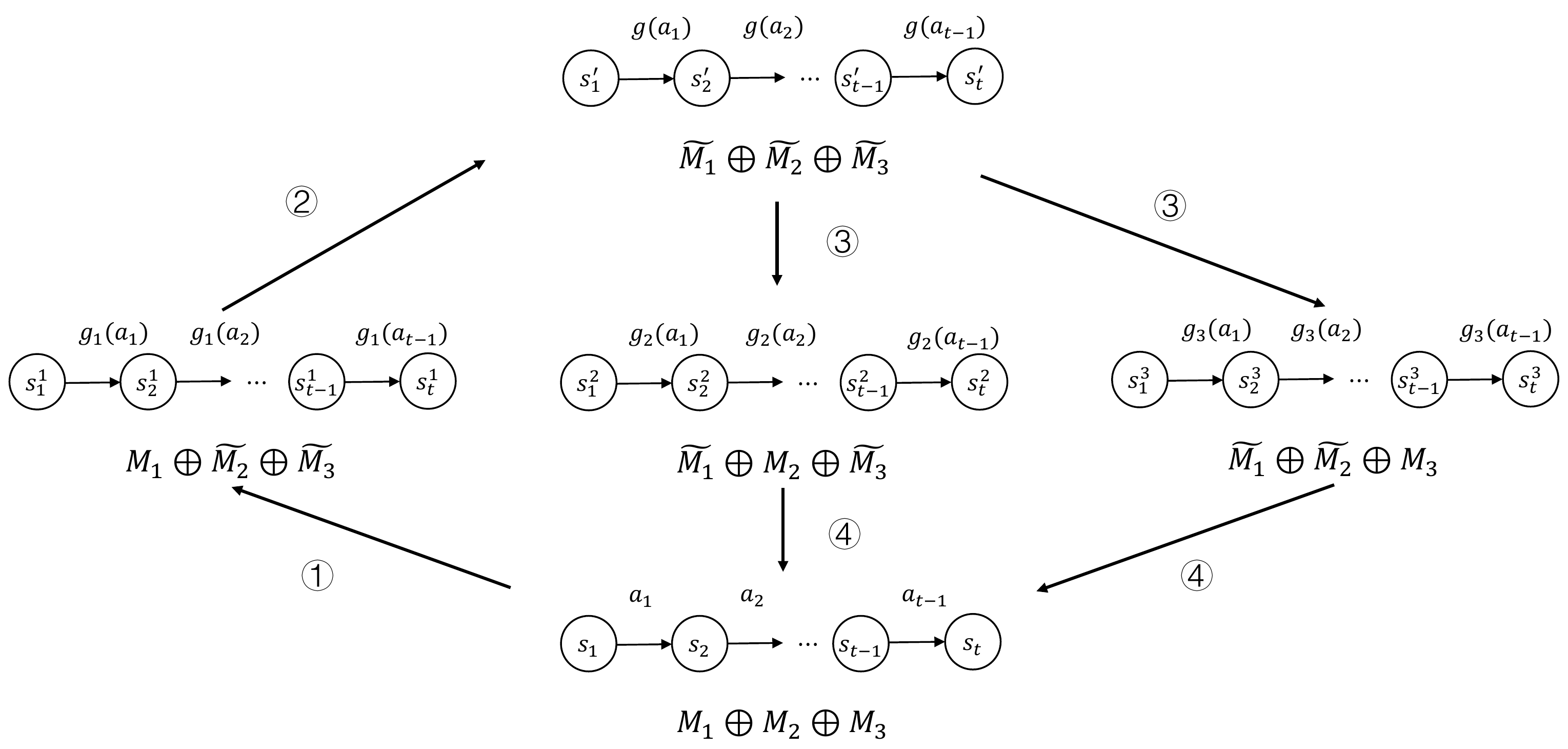}
    \caption{Correctness of Compositional Checking}
    \label{F: Chk}
\end{figure}

\noindent \textbf{Proof sketch.} Given the strong refinement mapping from each $C_i$ to $A$, for each valid trace of $S$ we construct a valid trace of $A$.
Both states and actions in the trace of $S$ are mapped to their counterparts in $A$, thus proving the strong refinement from $S$ to $A$. 
There are four steps in construction, as is shown in Figure \ref{F: Chk}. 
%--

\noindent \ding{172} Suppose $S$ takes an action $a\in M_i$ in current state. Then $C_i$ can also take $a$ because all actions in  $M_i$ are preserved by $C_i$. 

\noindent \ding{173} Given that $C_i\Rightarrow A$, we have the mapping $f$ of actions. Since $M_i$ is abstracted in $A$, $A$ can take the action $f(a)$. % to reach corresponding state. 

Note that in the current step we are considering an action $a$ from $M_i$. In the next step, $S$ may take an action from any module other than $M_i$.
Action $a$ in the current step may affect other modules which are taking an action in the next step.
So we have to ensure state of any compositional specification $C_j(j\neq i)$ remains consistent. We therefore need step \ding{174} and \ding{175}.

\noindent \ding{174} As module $M_i$ is abstracted in $C_j$, $C_j$ can take $f(a)$ just like $A$.

\noindent \ding{175} Comparing to $A$, module $M_j$ is not abstracted in $C_j$ and some internal variables of $M_j$ may be modified by $a$. We prove that $a$ and $f(a)$ modifies internal varianles of $M_j$ in the same way, thus ensuring the states of $S$ and $C_j$ remain consistent.

We provide the detailed proof in Appendix \ref{A: Proof}.

%--
%--
\section{Case Studies} \label{Sec: Case}

In this section, we apply the IPA framework to reduce the model checking cost for the specifications of two consensus protocols: Raft and PRaft.
Raft is a widely-used consensus protocol which is originally developed in the academia and then widely used in practice. 
PRaft is the replication protocol in PolarFS, the distributed file system for the commercial database Alibaba PoloarDB \cite{Cao18}. The design of PRaft is derived from Raft and Multi-Paxos \cite{Gu21}.

We first introduce the general pattern of interaction-preserving abstraction on realistic TLA+ specifications.
Then we demonstrate how the IPA framework can be conveniently applied in practice.
Finally, we show how much model checking cost can be saved through experiments.

Details on how each type of abstraction is conducted, including the TLA+ specifications before and after the abstraction, can be found in Appendix \ref{A: Case-Raft} and \ref{A: Case-PRaft}. All the TLA+ specifications in the Raft case can be found in the anonymized GitHub repository\footnote{https://github.com/AnonymousAccountForReview/IPA}. Up till now, TLA+ specifications in the PRaft case cannot be open-sourced due to confidentiality reasons.

%------------------------------
\subsection{Patterns of Interaction-Preserving Abstraction} \label{SubSec: Abs}

In Section \ref{Sec: Chk}, we present the constraints the abstraction must conform to, in order to guarantee interaction-preservation.
These constraints are necessary conditions and they do not tell the specification developer how to write the interaction-preserving abstractions in practice.
In this section, we show via case studies that the interaction-preserving abstraction is quite intuitive. Moreover, useful patterns can greatly mitigate the burden of the developer.

%--------------------
\subsubsection{The Polling Pattern}

Consensus protocols usually involve some polling process in one way or another, in order to collect local information from distributed nodes/replicas and calculate certain global information.
This type of polling process can generally be restricted within the scope of one function module. 
This means that other modules do not need to know the details of the polling process. They only care about the final result.
For example, the leader election module often needs to poll multiple candidates to choose the most eligible one. % which is most eligible to be the new leader. 
However, when we model check other modules, we only need to know which node is the new leader. 
% The leader election process can be and should be abstracted away.

\begin{figure}[htbp]
    \centering
    \includegraphics[width=0.8\linewidth]{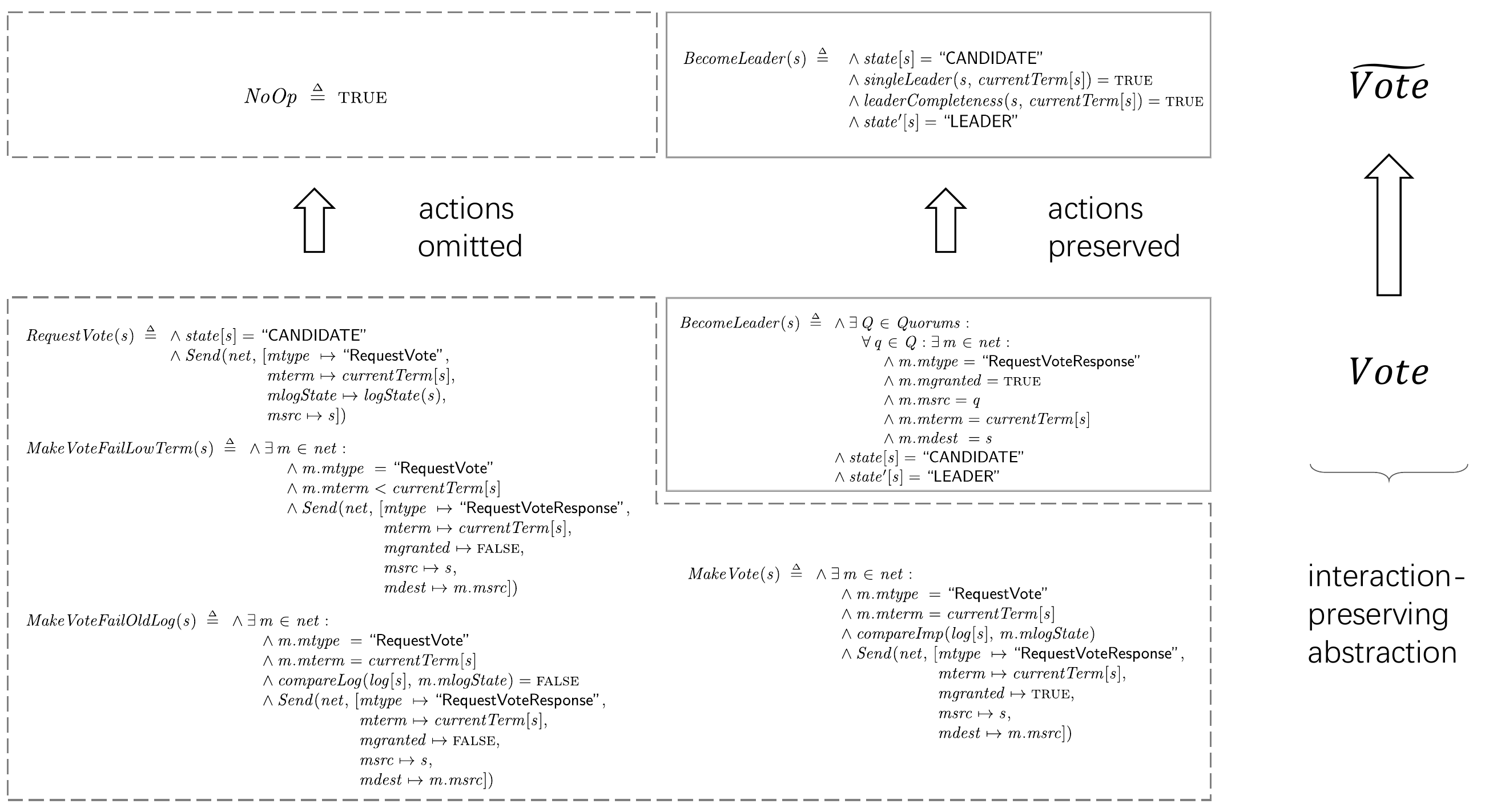}
    \caption{Interaction-preserving abstraction of the \textsf{Vote} module.}
    \label{F: Case}
\end{figure}

Thus, the details of the polling process can generally be abstracted away.
In TLA+, since specification developers model a distributed system in terms of a single global state, the abstraction is quite straightforward.
As shown in the illustrative example in Fig. \ref{F: Case}, utilizing the global information in the specification, the specification developer can obtain required global information in one step, without the polling process.
See more concrete examples of applying this ``polling" pattern in Appendix \ref{ASubSec: Replication}, \ref{ASubSec: PreVote}, \ref{ASubSec: Vote} and \ref{ASubSec: PRaft1}.
This abstraction process is intuitively correct, and we can conveniently double check its correctness following the constraints in Section \ref{Sec: Chk}.

% All other abstractions follow this pattern. See detailed description and specification in Appendix \ref{A: Case-Raft} and \ref{A: Case-PRaft}.

%--

%--------------------
\subsubsection{Industry Implementation Patterns}

In our case study, we intentionally choose the detailed specification of an industry-level consensus protocol PRaft.
To improve maintainability, reduce implementation complexity and support dynamic upgrade, PRaft separates its control flow from the data flow by introducing a centralized coordinator. The coordinator is in charge of the control plane and the leader node and follower nodes passively receive commands from the coordinator.
% which implements a set of sub protocols including leader election and follower recovery. In normal cases, servers receive and replicate commands from clients and execute those that reach consensus. 
The coordinator regularly checks each server's state to see whether any error occurs. When errors such as follower reboot or network partition are detected, the coordinator starts the corresponding error handling process by sending servers commands they need to execute. Servers are passive followers and never make decisions on their own.

A typical control flow of PRaft is as follows: 1) the coordinator sends a command to a server; 2) the server executes the command received; 3) the coordinator sends a message to the server requesting its progress; 4) the server responds telling the coordinator whether it has finished execution; 5) when the server finishes its current job, the coordinator sends the next command. Similar communications between the coordinator and the followers exist in both module \textsf{LeaderRecovery} and module \textsf{FollowerRecovery}.

Such control flows are suitable for system implementations but add unnecessary complexities to model checking. The relative order between these control flow communication steps and other actions are undetermined. When performing model checking, the coordinator is often redundant because a system specification stands at a global point of view and the specification developer can specify that servers make decisions on their own. % as if they receive requests from a virtual coordinator. 
Thus, step 1,3 and 5 can be omitted in abstraction. We apply this abstraction in multiple types of control flows of PRaft.

%\mybreak

Besides the abstraction concerning the control flow, we also find that industry-level design often pays a lot of attention to performance optimization in realistic scenarios.
From the perspective of interaction-preserving abstraction, the performance optimization protocol can often be replaced by a brute-force protocol. % This is indistinguishable from other modules, but can reduce the model checking cost.

For example, in PRaft, a centralized coordinator is introduced to instruct the operation of the leader node and the follower nodes. In the \textsf{LeaderRecovery} module, the coordinator has to calculate committed log entries from logs of a majority of servers. One way to implement this is that all nodes send their logs to the coordinator, taking one round of communication. But as the log is large, this may cause network congestion. To reduce the network load, PRaft uses two rounds of communications. In the first round, servers simply send the length of their logs to the coordinator, who selects a majority of servers whose log is more up-to-date. In the second round, only the selected server sends their logs to the coordinator. Thus network load is reduced using one more round of communication.
% Network capacity is not modeled in the TLA+ specification, so it has no effect on the cost of model checking. But more rounds of communication introduce more steps in behavior traces and more possible permutations of actions, which increase the cost of model checking. 
This two-round log collection protocol is replaced by the brute-force one-round protocol in the abstraction. %The two-round protocol is mainly for performance optimization and 

%The other IPA is just following the polling pattern. See details in Appendix \ref{A: Case-PRaft}.

%------------------------------
\subsection{Application of the IPA Framework}

After presenting how the interaction-preserving abstraction is conducted in principle, we now discuss important details in applying the IPA framework.

%--------------------
\subsubsection{Partitioning the TLA+ Specification into Modules}

Basically, the Raft specification can be divided into two modules: the \textsf{LogReplication} module, which describes how the leader replicates log entries to the followers, and the \textsf{Vote} module, which describes how a new leader is elected when the former leader fails.
In practice, the Raft protocol often includes the third module \textsf{PreVote}, which is used to prevent a disconnected follower from immoderately increasing the $term$ value. % and imposing unnecessary damage to the performance.

The TLA+ specification for PRaft is developed to precisely document its design, find potential deep bugs, and improve the developer's confidence in its design and implementation.
%PRaft is an industry protocol and our goal is to model PRaft close to implementation. 
%The specification preserves many industry-level optimizations. We conduct abstraction mainly by  omitting implementation details.
At the very beginning, the PRaft specification is divided into three modules: the \textsf{Replication} module, which describes how the leader replicates log entries to the followers, the \textsf{LeaderRecovery} module, which describes how a new leader is elected when the old leader fails, and the \textsf{FollowerRecovery} module, which describes how a lagged follower catches up. 
However, when we conduct interaction-preserving abstraction for each module, we find that the \textsf{Replication} module has little room for abstraction. 
It means that separating out this module will not reduce the overall compositional checking cost.
Therefore, we merge the \textsf{Replication} module into the \textsf{LeaderRecovery} module.

%--------------------
\subsubsection{Interaction-Preserving Abstraction in Practice}

Given the partition of TLA+ specification into modules, the key is to identify the interaction variables $\mc{I}$, thus identifying the internal variables $\mc{L}$.
In practical use of the IPA framework, the specification developer can easily classify the variables, since the developer is quite clear of the use of each variable when transforming the informal system design into TLA+ specifications. The high cohesion and low coupling of the modules also ease the burden of classifying the variables. 	
% During the process of describe the system design into TLA+, the use of each variable is clear. Understanding of the system logic can greatly help the developer classify the variables.
Given the intuitive and tentative classification of the variables, the developer only needs to double check the classification according to the constraints defined in Section \ref{SubSec: Module}.

The abstraction process basically follows intuitive patterns, as discussed in Section \ref{SubSec: Abs}.
Given that the developer has already transformed informal system design into detailed TLA+ specification, it is much easier to write the coarsened specification. 
During this process, the refinement mapping between the two levels of specifications is also intuitive.
Note that, the IPA framework requires strong refinement, while in TLC, we can only check (the original) refinement mapping. Currently, the strong refinement mapping is manually checked and guaranteed by introducing auxiliary variables.

%------------------------------
\subsection{Experimental Evaluation} \label{SubSec: Exp}

The main objective of the experimental evaluation is to explore how much model checking cost can be saved using our IPA framework.
We model check the Raft and PRaft specifications, and compare the cost in time between direct checking and compositional checking.
The model checking is conducted on one workstation with an Intel i9-9900X CPU (3.50GHz), with 10 cores and 20 threads, and 32GB RAM, running Ubuntu Desktop 16.04.6 LTS and TLC version 1.7.1.

We tune the scale of the system by tuning $term$ (the maximum number of phases the nodes can enter in the consensus process) and $cmd$ (the number of commands the clients can send to the servers).
%There are several parameters to consider for Raft such as the the maximum term servers can set and number of commands received from clients.
The number of servers is set to 3.
% On the one hand, we want to check how complexity change of a single module effect the cost of entire hybrid checking which demostrates the effectiveness of abstraction for that module. The maximun term servers can set and number of commands received from clients mainly affect complexity of module Vote and module Replication, respectively. 
%
% On the other hand, the rest parameters are assigned appropriate values so that the scenatios generated are non-trivial. We fix the number of servers to be three, which is typical for industrial system and is complex enough to expose possible errors in most cases. Similarly, clients can send two different commands to servers, which is the minimum number to find consistency violaitons and avoids unnecessary complexity.
%
We record the checking time for each module and obtain the overall time for compositional checking. We also record the time for direct checking.
The ratio of direct checking time to compositional checking time is calculated to illustrate the effect of compositional checking.

%--
\begin{table}[htbp]\tiny
	\caption{Experiment Results}
	\label{T: Result}
	\resizebox{\linewidth}{!}{
		\begin{tabular}{ccccccc}
			\toprule
			\multirow{2}{*}{\tabincell{c}{Raft \\ $(term,cmd)$}} & \multirow{2}{*}{$T_{\textsf{PreVote}}$} & \multirow{2}{*}{$T_{\textsf{Vote}}$}
			& \multirow{2}{*}{$T_{\textsf{Rep}}$} & \multirow{2}{*}{$T_{comp}$}
			& \multirow{2}{*}{$T_{direct}$} & \multirow{2}{*}{\tabincell{c}{$\frac{T_{direct}}{T_{comp}}$}} \\ 
			& &  &  &  &  & \\ \midrule
			(1,1)    &  00:00:06 & 00:00:05 & 00:00:03 & 00:00:14
			 & 00:02:25 & 10.3 \\ 
			(1,2)    &  00:00:14 & 00:00:14 & 00:00:06 & 00:00:34
			 & 01:03:48 & 111.2 \\ 
			(1,3)    &  00:01:27 & 00:01:38 & 00:00:57 & 00:04:02
			 & 19:12:50 & 288.6 \\ 
			(2,1)    &  00:00:38 & 00:00:14 & 00:00:09 & 00:01:01 & 03:27:06 & 203.7 \\ 
			(2,2)    &  00:08:37 & 00:03:08 & 00:09:54 & 00:21:39 & $>$100:00:00 & $>$277.1 \\ 
			(2,3)    &  05:20:22 & 01:00:11 & 39:57:55 & 46:18:23 & $>$200:00:00 & $>$4.3 \\ 
			\midrule
			\midrule
			\multirow{2}{*}{\tabincell{c}{PRaft \\ $(term,cmd)$}} & \multicolumn{3}{c}{\multirow{2}{*}{$T_{\textsf{RecL}}$} \qquad \multirow{2}{*}{$T_{\textsf{RecF}}$}}   &  \multirow{2}{*}{$T_{comp}$} & 
			\multirow{2}{*}{$T_{direct}$} & \multirow{2}{*}{\tabincell{c}{$\frac{T_{direct}}{T_{comp}}$}} \\ % {Reduction \\Ratio}} \\ 
			& &  &  &  &  & \\ \midrule
			(1,1)    &  \multicolumn{3}{c}{00:00:04 \qquad 00:00:14}   & 00:00:18
			 & 00:00:35 & 1.9 \\ 
			(1,2)    &  \multicolumn{3}{c}{00:01:34 \qquad 00:05:44}  & 00:07:18
			 & 00:21:10 & 2.9 \\ 
			(1,3)    &  \multicolumn{3}{c}{00:23:05 \qquad 04:30:50}   & 04:53:55 & 13:54:05 & 2.8 \\ 
			(2,1)    & \multicolumn{3}{c}{02:47:43 \qquad 00:37:56}   & 03:25:39
			 & 09:48:20 & 2.9 \\ 
			(2,2)    & \multicolumn{3}{c}{33:24:07 \qquad 31:58:08}   & 65:22:15
			 & $>$200:00:00 & $>$3.1 \\ 
			\bottomrule
			
		\end{tabular}
	}
\end{table}

The experiment results are listed in Table \ref{T: Result}. $T_{\textsf{PreVote}}$ denotes the compositional model checking of module \textsf{PreVote} and the checking time of other modules are named similarly. $T_{comp}$ denotes the total compositional checking time for all modules and $T_{direct}$ denotes the time for direct checking of the original specification. In our analysis of the evaluation results, we mainly investigate the cost ratio, which is defined as $\frac{T_{direct}}{T_{comp}}$.

As for the Raft case, the cost ratio ranges from 10.3 to 288.6, showing that compositional checking based on the IPA framework can significantly reduce the model checking cost. Principally, the more complicated the model is, the larger the cost ratio. This is mainly because for complex modules, there will be more internal logic which can be abstracted away in the compositional checking.
Note that in the case where $(term, cmd) = (2,3)$, we stop the direct checking when the total checking time reaches 200 hours. So the result that $cost\ ratio > 4.3$ is a quite conservative estimation. It is reasonable to estimate that the actual cost ratio is much more than 4.3, probably also much more than 288.6.

As for the PRaft case, the cost ratio is around 3, relatively small compared to the ratio in the Raft case.
It is mainly because, although the PRaft protocol is derived from Raft, it works much more like Multi-Paxos. Thus the abstractions in the Raft case are not applicable in the PRaft case.
Moreover, in the PRaft case, we mainly abstract away the details of performance optimizations. Such details consist a smaller portion in the protocol design, compared to the Raft case.
Although the cost ratio is smaller in the PRaft case, we argue that the IPA framework is practically effective in the PRaft case. It can save much time compared to the direct checking. Also note that, in the PRaft case, the TLA+ specifications are supplemented after the protocol design and implementation are principally finished, in order to precisely document the protocol design and find potential deep bugs in the implementation. Thus the abstraction process is intuitive and in some sense straightforward, for developers who are familiar with the PRaft design. This makes the application of the IPA framework highly worthwhile.

%--
%--
\section{Related Work} \label{Sec: RW}

Compositional model checking is essential to tackling the state explosion problem.
It can be roughly classified as compositional minimization and compositional reasoning \cite{Zheng10}. 
In compositional reasoning, verification of a system is broken into separate analyses for each component of the system. 
The result for the entire system is derived from the results of verifying individual components \cite{Clarke89,Pnueli89,Grumberg94,Henzinger98}. 
In our approach, after the abstraction of each module is obtained, the following compositional checking is fully automatic.
%detailed specification, the abstraction is by hand, but is much more straightforward.
%After the abstraction, our approach is just mode checking. Our model checking does not impose such burden.
The compositional reasoning imposes non-trivial burden on the developer, and it is not suitable for the intended users of our IPA framework. 

%The success of compositional reasoning relies on discovery of appropriate environment assumptions for every component.
%If the components have complex interactions with their environments, generating accurate environment assumptions can be challenging. Therefore, the requirement of manually finding assumptions has been a factor limiting the practical use of compositional reasoning.

Compositional minimization, in general, constructs the local model for each module in a system, minimizes it, and composes it with the minimized models of other modules to form a reduced global model for the entire system, on which verification is performed \cite{Graf96,Krimm97}.
Effectiveness of these methods depends on whether a coarse enough (to reduce the checking cost) yet accurate enough (to ensure the correctness of checking) context can be found for each component such that all the essential behavior of that component can be checked.
% Refinement of environment assumptions in compositional verification [1], [20], [28].
% In \cite{Alfaro01}, the interface automata is proposed to represent a module and its environment. The module and the environment are refined in an alternating fashion so that the module accepts only input actions generated by the environment, and issues output actions corresponding to these input actions.
Existing compositional minimization techniques do not consider the characteristics of TLA+ specifications, and are thus not applicable or efficient in our target scenarios.
Our IPA framework achieves compositional minimization based on the ladder of abstractions in TLA+ specifications. 
%Specifically, in TLA+, correctness properties and system designs are just steps on a ladder of abstraction, with correctness properties occupying higher levels, systems designs and algorithms in the middle, and executable code and hardware at the lower levels.
%The designer can then verify that each level is correct with respect to a higher level.
The freedom to choose and adjust levels of abstraction is utilized to achieve the compositional minimization we need. %The main challenge is to achieve interaction-preservation.

% Developers utilize the ladder of abstraction to manage the complexity of real-world systems. 
% Developers may choose to describe the system at several “middle” levels of abstraction, with each lower level serving a different purpose (such as to understand the consequences of finer-grain concurrency or more detailed behavior of a communication medium). 

The interaction-preservation abstraction of this work is also inspired by the dynamic interface reduction technique in code-level model checking \cite{Guo11}.
The dynamic interface reduction technique essentially identifies the interface interactions between running nodes of a distributed system and eliminates traces with the same interface behaviors so that the state space to be checked is reduced.
Our compositional minimization is orthogonal to the reduction of model checking state space, but we borrow the basic idea of interface reduction.

\section{Conclusion and Future Work} \label{Sec: Concl}

In this work we present the IPA compositional model checking framework for TLA+ specifications of consensus protocols.
%which closes the gap between classical modular checking algorithms and practical TLA+ applications. 
We provide formal definition and correctness proof of our IPA framework. 
We also apply the IPA framework in model checking of two consensus protocols Raft and PRaft.
The case study shows that the IPA framework is easy to use in practical model checking of realistic TLA+ specifications.
It also shows that the compositional model checking based on IPA can significantly reduce the checking cost.
% We identify two general patterns to apply IPA and conduct experiments in two cases,  showing that IPA is applicable in both acadamic and industrial projects. 

In our future work, we will apply the IPA framework to more scenarios, involving complex and subtle distributed protocols other than distributed consensus.
We will also investigate whether the IPA framework can be used to reduce the cost of code-level model checking of distributed system implementations.
Given sufficient application of the IPA framework in realistic scenarios, we will investigate how to integrate the IPA framework into the extreme modeling \cite{Davis20} paradigm of distributed system design and implementation.

%-- bib
\bibliographystyle{splncs04}
\bibliography{IPA-CAV22}

\begin{thebibliography}{10}
\providecommand{\url}[1]{\texttt{#1}}
\providecommand{\urlprefix}{URL }
\providecommand{\doi}[1]{https://doi.org/#1}

\bibitem{Etcd}
https://etcd.io/

\bibitem{TLA}
https://lamport.azurewebsites.net/tla/tla.html

\bibitem{Paxos-TLA}
https://github.com/tlaplus/DrTLAPlus/tree/master/Paxos

\bibitem{Raft-TLA}
https://github.com/ongardie/raft.tla

\bibitem{ZK-TLA}
https://github.com/apache/zookeeper/pull/1690

\bibitem{Bornholt21}
Bornholt, J., Joshi, R., Astrauskas, V., Cully, B., Kragl, B., Markle, S.,
  Sauri, K., Schleit, D., Slatton, G., Tasiran, S., Van~Geffen, J., Warfield,
  A.: Using lightweight formal methods to validate a key-value storage node in
  amazon s3. In: Proceedings of the ACM SIGOPS 28th Symposium on Operating
  Systems Principles. p. 836–850. SOSP '21, Association for Computing
  Machinery, New York, NY, USA (2021). \doi{10.1145/3477132.3483540},
  \url{https://doi.org/10.1145/3477132.3483540}

\bibitem{Burrows06}
Burrows, M.: The {Chubby} lock service for loosely-coupled distributed systems.
  In: Proc. OSDI'06, USENIX Symposium on Operating Systems Design and
  Implementation. pp. 335--350. USENIX (2006),
  \url{http://dl.acm.org/citation.cfm?id=1298455.1298487}

\bibitem{Cao18}
Cao, W., Liu, Z., Wang, P., Chen, S., Zhu, C., Zheng, S., Wang, Y., Ma, G.:
  Polarfs: An ultra-low latency and failure resilient distributed file system
  for shared storage cloud database. Proc. VLDB Endow.  \textbf{11}(12),
  1849--1862 (aug 2018). \doi{10.14778/3229863.3229872},
  \url{https://doi.org/10.14778/3229863.3229872}

\bibitem{Chandra07}
Chandra, T.D., Griesemer, R., Redstone, J.: Paxos made live: An engineering
  perspective. In: Proceedings of the Twenty-sixth Annual ACM Symposium on
  Principles of Distributed Computing. pp. 398--407. PODC '07, ACM (2007),
  \url{http://doi.acm.org/10.1145/1281100.1281103}

\bibitem{Clarke89}
{Clarke}, E.M., {Long}, D.E., {McMillan}, K.L.: Compositional model checking.
  In: [1989] Proceedings. Fourth Annual Symposium on Logic in Computer Science.
  pp. 353--362 (1989). \doi{10.1109/LICS.1989.39190}

\bibitem{Clarke00}
Clarke, E.M., Grumberg, O., Peled, D.A.: Model Checking. MIT Press, Cambridge,
  MA, USA (2000)

\bibitem{Corbett12}
Corbett, J.C., Dean, J., Epstein, M., Fikes, A., Frost, C., Furman, J.J.,
  Ghemawat, S., Gubarev, A., Heiser, C., Hochschild, P., Hsieh, W., Kanthak,
  S., Kogan, E., Li, H., Lloyd, A., Melnik, S., Mwaura, D., Nagle, D., Quinlan,
  S., Rao, R., Rolig, L., Saito, Y., Szymaniak, M., Taylor, C., Wang, R.,
  Woodford, D.: Spanner: {Google's} globally-distributed database. In: Proc.
  OSDI'12, USENIX Symposium on Operating Systems Design and Implementation. pp.
  251--264. USENIX (2012),
  \url{http://dl.acm.org/citation.cfm?id=2387880.2387905}

\bibitem{Davis20}
Davis, A.J.J., Hirschhorn, M., Schvimer, J.: Extreme modelling in practice.
  Proc. VLDB Endow.  \textbf{13}(9),  1346--1358 (May 2020).
  \doi{10.14778/3397230.3397233},
  \url{https://doi.org/10.14778/3397230.3397233}

\bibitem{Graf96}
Graf, S., Steffen, B., L{\"u}ttgen, G.: Compositional minimisation of finite
  state systems using interface specifications. Formal Aspects of Computing
  \textbf{8}(5),  607--616 (1996)

\bibitem{Grumberg94}
Grumberg, O., Long, D.E.: Model checking and modular verification. ACM Trans.
  Program. Lang. Syst.  \textbf{16}(3),  843–871 (may 1994).
  \doi{10.1145/177492.177725}, \url{https://doi.org/10.1145/177492.177725}

\bibitem{Gu21}
Gu, X., Wei, H., Qiao, L., Huang, Y.: Raft with out-of-order executions.
  International Journal of Software and Informatics  \textbf{11}(4), ~473
  (2021). \doi{10.21655/ijsi.1673-7288.00257}

\bibitem{Paz18}
Guay~Paz, J.R.: Microsoft Azure Cosmos DB Revealed: A Multi-Model Database
  Designed for the Cloud. Apress, Berkeley, CA (2018)

\bibitem{Guo11}
Guo, H., Wu, M., Zhou, L., Hu, G., Yang, J., Zhang, L.: Practical software
  model checking via dynamic interface reduction. In: Proceedings of the
  Twenty-Third ACM Symposium on Operating Systems Principles. pp. 265--278.
  SOSP '11, ACM, New York, NY, USA (2011). \doi{10.1145/2043556.2043582},
  \url{http://doi.acm.org/10.1145/2043556.2043582}

\bibitem{Henzinger98}
Henzinger, T.A., Qadeer, S., Rajamani, S.K.: You assume, we guarantee:
  Methodology and case studies. In: Hu, A.J., Vardi, M.Y. (eds.) Computer Aided
  Verification. pp. 440--451. Springer Berlin Heidelberg, Berlin, Heidelberg
  (1998)

\bibitem{Hunt10}
Hunt, P., Konar, M., Junqueira, F.P., Reed, B.: {ZooKeeper}: wait-free
  coordination for internet-scale systems. In: Proc. ATC'10, USENIX Annual
  Technical Conference. pp. 145--158. USENIX (2010),
  \url{http://portal.acm.org/citation.cfm?id=1855840.1855851}

\bibitem{Junqueira11}
Junqueira, F.P., Reed, B.C., Serafini, M.: Zab: high-performance broadcast for
  primary-backup systems. In: Proc. DSN'11, IEEE/IFIP Conference on Dependable
  Systems and Networks. pp. 245--256. IEEE (2011),
  \url{http://dx.doi.org/10.1109/DSN.2011.5958223}

\bibitem{Krimm97}
Krimm, J.P., Mounier, L.: Compositional state space generation from lotos
  programs. In: Brinksma, E. (ed.) Tools and Algorithms for the Construction
  and Analysis of Systems. pp. 239--258. Springer Berlin Heidelberg, Berlin,
  Heidelberg (1997)

\bibitem{Lamport01}
Lamport, L.: Paxos made simple. ACM SIGACT News (Distributed Computing Column)
  32, 4 (Whole Number 121, December 2001) pp. 51--58 (December 2001),
  \url{https://www.microsoft.com/en-us/research/publication/paxos-made-simple/}

\bibitem{Leesatapornwongsa14}
Leesatapornwongsa, T., Hao, M., Joshi, P., Lukman, J.F., Gunawi, H.S.: Samc:
  Semantic-aware model checking for fast discovery of deep bugs in cloud
  systems. In: Proceedings of the 11th USENIX Conference on Operating Systems
  Design and Implementation. pp. 399--414. OSDI'14, USENIX Association,
  Berkeley, CA, USA (2014),
  \url{http://dl.acm.org/citation.cfm?id=2685048.2685080}

\bibitem{Newcombe15}
Newcombe, C., Rath, T., Zhang, F., Munteanu, B., Brooker, M., Deardeuff, M.:
  How amazon web services uses formal methods. Commun. ACM  \textbf{58}(4),
  66--73 (Mar 2015). \doi{10.1145/2699417},
  \url{https://doi.org/10.1145/2699417}

\bibitem{Ongaro14}
Ongaro, D., Ousterhout, J.: In search of an understandable consensus algorithm.
  In: Proceedings of the 2014 USENIX Conference on USENIX Annual Technical
  Conference. pp. 305--320. USENIX ATC'14, USENIX Association, Berkeley, CA,
  USA (2014), \url{http://dl.acm.org/citation.cfm?id=2643634.2643666}

\bibitem{Pnueli89}
Pnueli, A.: In Transition from Global to Modular Temporal Reasoning about
  Programs, p. 123–144. Springer-Verlag, Berlin, Heidelberg (1989)

\bibitem{Zheng10}
Zheng, H., Yao, H., Yoneda, T.: Modular model checking of large asynchronous
  designs with efficient abstraction refinement. IEEE Transactions on Computers
   \textbf{59}(4),  561--573 (2010)

\end{thebibliography}

%--
\newpage

%--
%--
\appendix

%--------------------------------------------------------------------------------
\section{Correctness Proof of the Compositional Checking} \label{A: Proof}

We here present the detailed proof of Theorem \ref{Th: Chk} in Section \ref{SubSec: Proof}.

\noindent \textbf{Proof.} We prove by mathematical deduction that for all valid traces of $S$ and $1\leq i\leq k$, the traces deduced using mapping from actions of $S$ to that of $C_i$ are valid traces of $C_i$ and that variables in $V_{C_i}$ have the same values in corresponding states. First we assume in all specifications, the same variables are assigned the same values in the initial states. Suppose the proposition holds for all traces whose length are smaller than $t$.

Let $s_1\stackrel{a_1}{\rightarrow}s_2\stackrel{a_2}{\rightarrow}\cdots s_{t-1}\stackrel{a_{t-1}}{\rightarrow}s_t$ be any valid trace of $S$ whose length is $t$. By induction hypothesis, we know that for any $1\leq i\leq k$, $s_{1}^i \stackrel{g_i(a_1)}{\rightarrow}s_{2}^i \stackrel{g_i(a_2)}{\rightarrow} \cdots \stackrel{g_{i(a_{t-2})}}{\rightarrow}s_{t-1}^i$ is a valid trace of $C_i$ and variables in $V_{C_i}$ are assigned the same values in $s_l$ and $s_{l}^i(1\leq l\leq t-1)$. Because $C_i\Rightarrow A$, accroding to the definition of $g$ and $\bar{g}_i$, we get that $s_1'\stackrel{g(a_1)}{\rightarrow} s_2'\stackrel{g(a_2)}{\rightarrow}\cdots \stackrel{g(a_{t-2})}{\rightarrow}s_{t-1}'$ is a valid trace of $A$ and also variables in $V_A$ are assigned the same values in $s_l'$ as in $s_l(1\leq l\leq t-1)$. 

Suppose $S$ executes an action $a_{t-1} \in M_i$ in state $s_{t-1}$ and reach state $s_t$.  As variables in $V_{C_i}$ are assigned the same values in $s_{t-1}$ as in $s_{t-1}^i$ and $D_i \subseteq V_{C_i}$, every enabling condition of $a_{t-1}$ must be satisfied in state $s_{t-1}^i$. Therefore, $S_i$ can also execute $a_{t-1}$ in state $s_{t-1}^i$ to reach state $s_{t}^i$ with variales in $V_{C_i}$ still are assigned the same values in $s_t$ as in $s_{t}^i$. 

Since $C_i \Rightarrow A$, $A$ can execute $g(a_{t-1})$ in state $s_{t-1}'$ to reach state $s_t'$ with variables in $V_A$ are assigned the same values in $s_t$ as in $s_t'$.

For any specificaion $S_j(j\neq i)$, as variables in $V_A$ have the same values in $s_{t-1}^j$ as in $s_{t-1}'$, $S_j$ can also execute $g_j(a_{t-1})=g(a_{t-1})$ to reach state $s_{t}^j$ with $V_A$ still assigned the same values in $s_t'$ and $s_t$ as in $s_{t}^j$.

For the internal variables $\mathcal{L}_{M_j} = \mathcal{D}_{M_j}\setminus \mathcal{I}$ of $M_j$, according to the definition of $f_i$, $a_{t-1}$ has exactly the same assignment clauses to $\mathcal{L}_{M_j}$ as $g_j(a_{t-1})$ and variables in $V_{C_j}$ have the same values in $s_{t-1}$ as in $s_{t-1}^j$, so variables in $\mathcal{L}_{M_j}$ are assigned the same values by $a_{t-1}$ and $g_j(a_{t-1})$. Thus, the values of variables in $V_{C_j}$ are the same in $s_{t}^j$ and $s_t$. 

As the trace is chosen arbitrarily, by mathematical induction, we get that for any $1\leq i\leq k$, $S\Rightarrow C_i(1\leq i\leq k)$. As $C_i\Rightarrow A$, we get $S\Rightarrow A$.

%--------------------------------------------------------------------------------
\section{Case Study on Raft} \label{A: Case-Raft}

We divide the specification of the Raft protocol into three modules: module \textsf{PreVote} describing the pre-vote mechanism, module \textsf{Vote} describing the election mechanism and module \textsf{Replication} describing the transmission of log entries from the leader to the followers. For each module, we mainly discuss how the interaction-preserving abstraction is conducted, as detailed below.

%----------------------------------------
\subsection{Abstraction for Module Replication} \label{ASubSec: Replication}

%--
\begin{figure}[htbp]
	\centering
	\includegraphics[width=\linewidth]{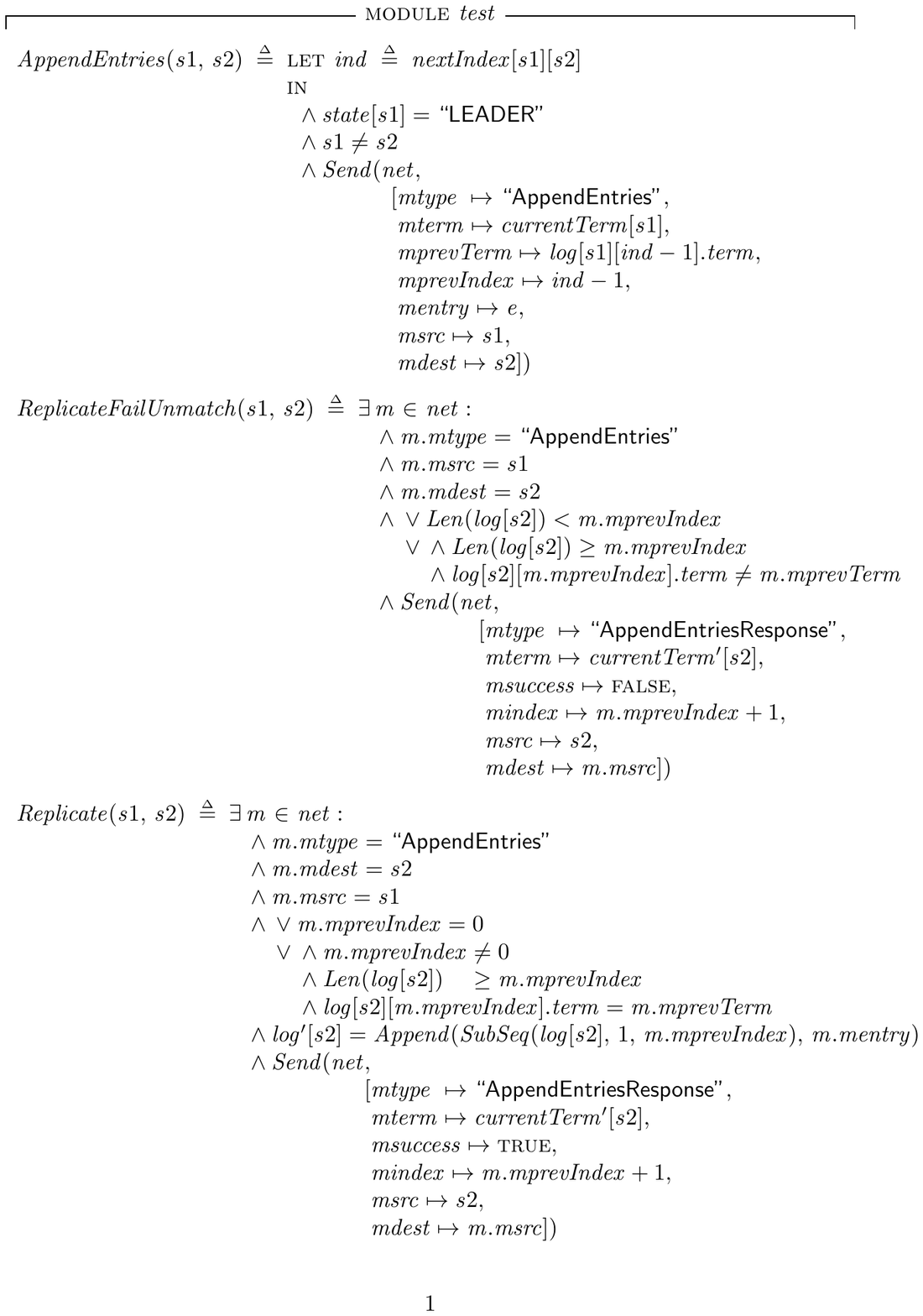}
	\caption{Specification for \textsf{Replication}}
	\label{F:ReplicationA}
\end{figure}

%--
\begin{figure}[htbp]
	\centering
	\includegraphics[width=\linewidth]{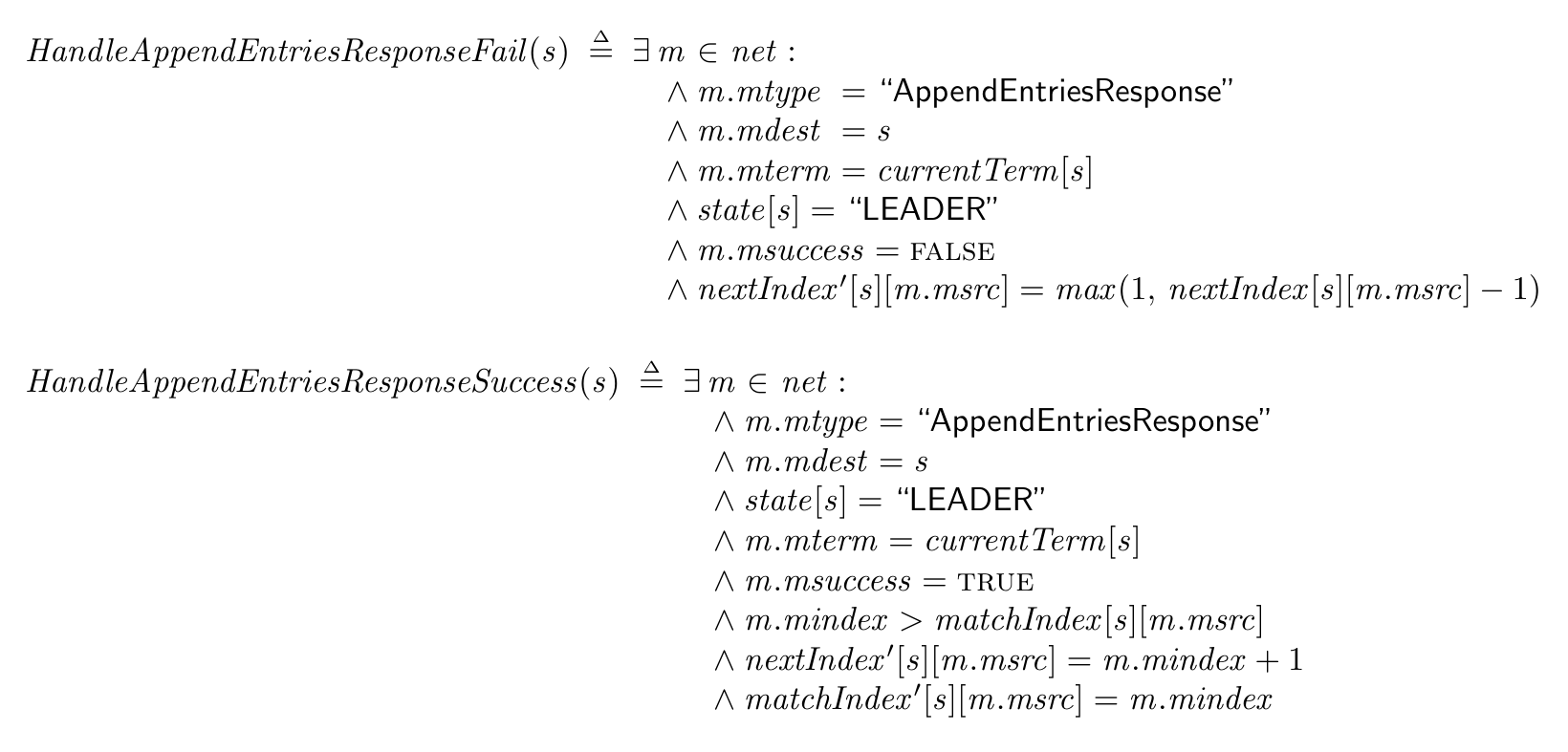}
	\caption{Specification for \textsf{Replication} (continued)}
	\label{F:ReplicationB}
\end{figure}

In Raft, log entries are replicated from leader to follower in sequence. A follower accepts a log entry from leader only if it has already accepted all previous log entries. If it receives a log entry when preceding entries are not fully accepted, it rejects the entry. When an entry is rejected, leader tries to send the previous one in its log. As leader does not know exactly which entry the follower would accept in an asynchronous distributed system, the process of sending and rejecting may take multiple rounds before the follower can accept its missing entry, in which the state variables of the leader and follower remain unchanged. Any newly elected leader has to find each follower's first unmatched log entry by such process.
Therefore, a trace containing multiple elections can be very long due to such ``invalid'' communications between leader and followers. Also many system states are generated due to the uncertain order between these invalid actions and other actions.

Figure \ref{F:ReplicationA} lists three actions of module \textsf{Replication}. Action $AppenEntries$ specifies the process when leader $s_1$ sends an $AppendEntries$ request to some follower $s_2$ to replicate log entries within the cluster. This action only modifies internal variable $net$ which records all messages sent by servers. When a follower receives a log entry from leader, it performs the prefix check to ensure that it has already received all previous log entries. If prefix check fails, the follower simply responds to leader with the index of unmatched entry without modifying any other variable, as is specified by action $ReplicateFailUnmatch$. If a follower receives from leader exactly the log entry it misses, it adds the entry to its log and sends back an ack, as is specified by $Replicate$. Figure \ref{F:ReplicationB} lists actions specifying how leader handles responses from followers. When leader receives a response from a follower indicating that the entry is accepted, it records the match index and the index of next log entry to send to the follower, as is specified by $HandleAppendEntriesResponseSuccess$. Note that $matchIndex$ and $nextIndex$ are specific to replication mechanism implementation and thus internal variables which can be omitted in the abstracted specification. When a follower rejects the log entry from leader, leader learns that this log entry is not the first one the follower misses. So it reduces $nextIndex$ for the follower by 1 and tries to replicate the last log entry. Action $HandleAppendEntriesResponseFail$ specifies this process.

%--
\begin{figure}[htbp]
	\centering
	\includegraphics[width=\linewidth]{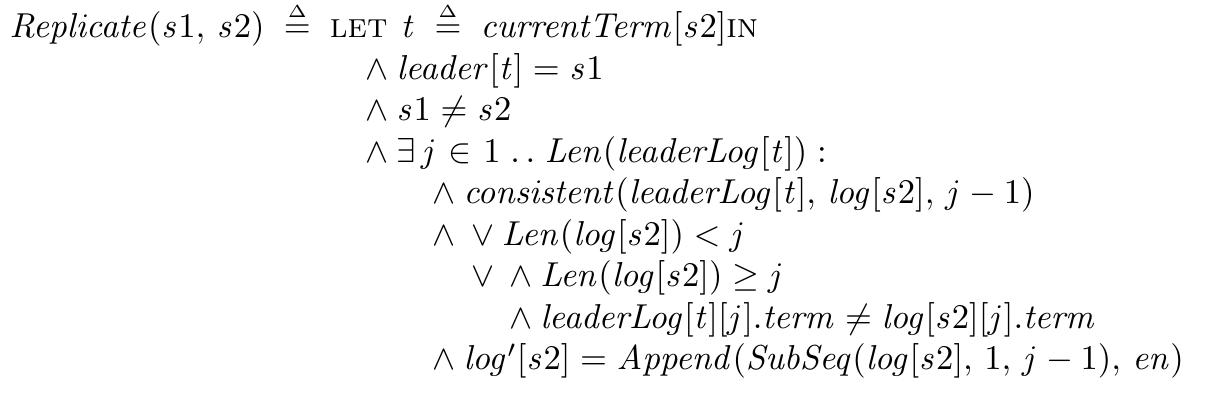}
	\caption{Abstracted Specification for \textsf{Replication}}
	\label{F:AbsReplication}
\end{figure}

To conduct abstraction, a leader simply sends the exact log entry that followers miss without the process of trial and error because TLA+ allows users to model a distributed system in terms of a single global state and a leader can utilize global state of each server. Thus, the redundant steps of sending and receiving ``invalid''  messages in system traces are eliminated. 
Figure \ref{F:AbsReplication} shows the single action of the abstracted specification for this process. This action corresponds to action $Replicate$ in Figure \ref{F:ReplicationA}. All other actions and internal variables such as $nextIndex$ and $matchIndex$ are reduced by abstraction.

%----------------------------------------
\subsection{Abstraction for Module PreVote} \label{ASubSec: PreVote}

Raft relies on a leader election algorithm to elect a single leader for each term. If follower does not receive heart beat messages from leader for some time, it becomes candidate and starts an election by increasing its term and sending election messages concurrently to other system servers. Since network may be unreliable, a server partitioned from leader cannot receive messages from leader. Therefore, it tries to start election for multiple times and increases its term to a large value. When network condition becomes normal, its large term would be propagated within the cluster, forcing the leader to step down and the cluster has to elect a new leader unnecessarily.

To prevent such occasional network fluctuation from causing disruptions, Raft introduces the pre-vote mechanism. When a follower tries to start an election, it has to send pre-vote requests to other servers. Servers grant or refuse pre-vote requests based on their system states. Only if the server learns from a majority of the cluster that they would grant its pre-vote request can it increases its term and make election proposals. Pre-vote mechanism solves the issue of partitioned server disrupting the cluster when it rejoins since a partitioned server cannot increase its term unless a majority of the cluster agree to elect a new leader.

\begin{figure}[htbp]
	\centering
	\includegraphics[width=\linewidth]{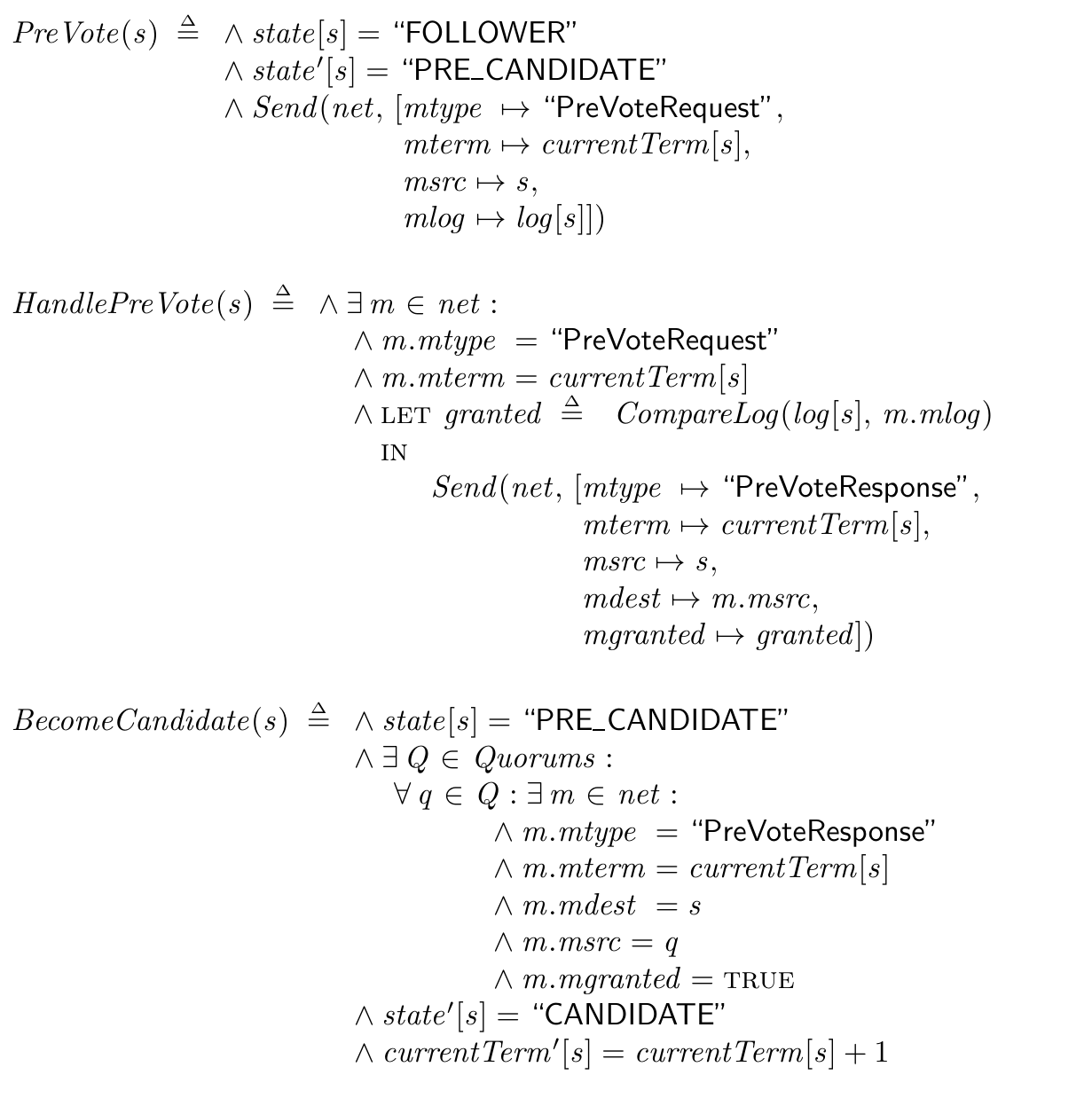}
	\caption{Specification for \textsf{PreVote}}
	\label{F:PreVote}
\end{figure}

\begin{figure}[htbp]
	\centering
	\includegraphics[width=\linewidth]{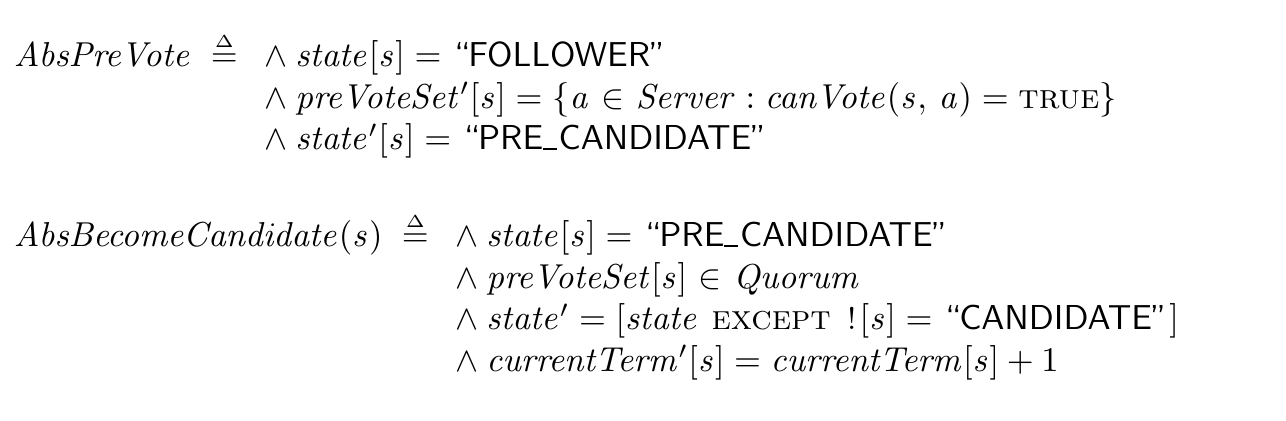}
	\caption{Abstracted Specification for \textsf{PreVote}}
	\label{F:AbsPreVote}
\end{figure}

Figure \ref{F:PreVote} shows the three actions of module \textsf{PreVote}. The behavior that a follower starts an election by changing its state to $PreCandidate$ and sending election requests to other servers for network reasons is modeled by action $PreVote$. Servers handle received election requests by executing action $HandlePreVote$. If the sender receives granted responses from a majority of servers, it changes state to $Candidate$ and updates its term, as is specified by action $BecomeCandidate$.

Note that according to action $HandlePreVote$, servers handle pre-vote requests by sending response messages without modifying their system states, which means that the action is transparent to other modules. We can thus omit this action in the abstraction of module \textsf{PreVote}. Figure \ref{F:AbsPreVote} is the abstracted version containing 2 actions, each corresponding to $PreVote$ and $BecomeCandidate$ respectively. In action $AbsPreVote$, except for changing follower's state, a history variable $preVoteSet$ is used to record all possible servers that may grant the follower's pre-vote request. The follower trying to start an election can change state to $Candidate$ only if its $preVoteSet$ contains a quorum of servers, as is specified by action $AbsPreVote$. In this way, the action of handling election request is omitted while preserving the functionality of pre-vote mechanism.

%----------------------------------------
\subsection{Abstraction for Module Vote} \label{ASubSec: Vote}

Raft's leader election algorithm can also be abstracted similarly. To become a leader, a candidate sends election request currently to all servers requesting for votes. A server may grant or refuse a vote request according to its state and the information the message contains. Only when the candidate's vote is approved by a majority of servers can it become leader. Thus, a round of election takes multiple steps in a behavior trace as servers work asynchronously and each server's handling of the vote request takes one step. More over, as network is unreliable and messages can be delayed arbitrarily, the order each server handles the election request is undetermined, adding much more system states to be checked.

A natural abstraction for election is to choose a server and change its role to leader, taking only one step and avoiding possible permutations due to asynchrony. However, the election algorithm is delicately designed to ensure that every newly elected leader meets several essential properties such as single leader and leader completeness, which are critical to the correctness of Raft. Therefore, we figure out and specify these properties in our specification. With this, we can specify that an eligible server change its state to leader as abstraction for election, omitting details while perfectly matching original design. 

\begin{figure}[htbp]
	\centering
	\includegraphics[width=\linewidth]{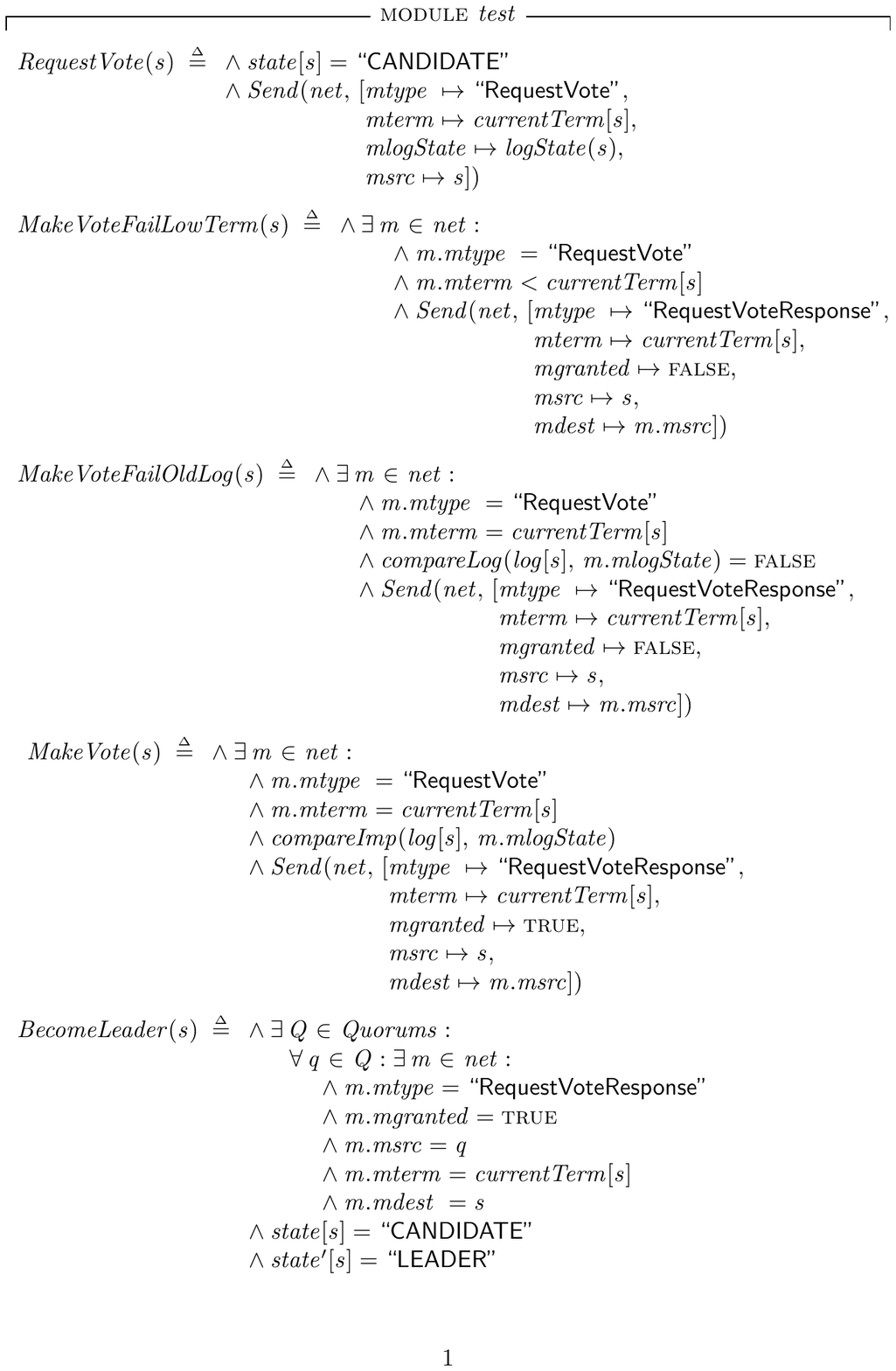}
	\caption{Specification for \textsf{Vote}}
	\label{F:Vote}
\end{figure}

Figure \ref{F:Vote} shows the four actions of module \textsf{Election}. Action $RequestVote$ specifies a candidate sends election requests to all other servers when starting a new election. This action only changes variable $net$, which records all messages sent by servers. $net$ is an internal variable, so action $RequestVote$ can be omitted by abstraction. Servers may grant or refuse election requests by comparing candidate's $term$ and $log$ with their own. Action $MakeVoteFailLowTerm$ specifies the case when a server refuses an election request because the candidate's $term$ is smaller. Action $MakeVoteFailOldLog$ specifies the case when the candidate's election request is refused because of outdated log. Both these two actions change no variables except for $net$ and thus can also be omitted. If the candidate's $term$ and $log$ are no older than the follower who received candidate's election reuqest, it grants the election request by sending an ack, as is specified by action $MakeVote$. This action also only modifies variable $net$. If a candidate receives ack from a quorum of servers, it changes its state to $LEADER$, as action $BecomeLeader$ specifies.

\begin{figure}[htbp]
	\centering
	\includegraphics[width=\linewidth]{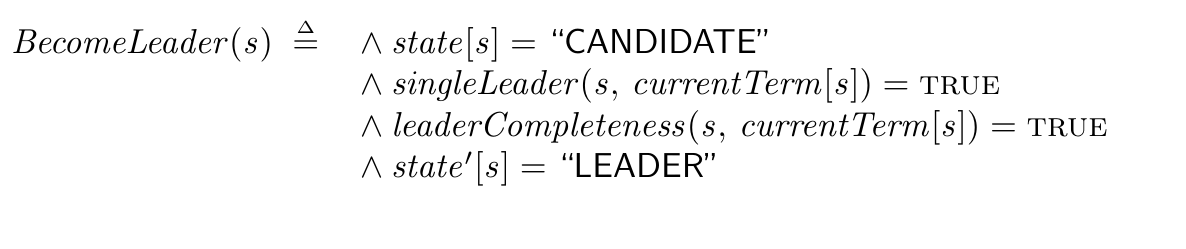}
	\caption{Abstracted Specification for \textsf{Vote}}
	\label{F:AbsVote}
\end{figure}

Figure \ref{F:AbsVote} shows the abstracted specification of leader election. It contains only one action, $BecomeLeader$, which is enabled only if the two essential properties $singleLeader$ and $leaderCompleteness$, are true. Thus, any successful leader election in the abstracted specification guarantee these two properties. By such abstraction, a round of election takes only 2 steps no matter how many servers are in the cluster, greatly reducing the complexity of election algorithm, especially when the number of servers is big.

%------------------------------
\section{Case Study on ParallelRaft} \label{A: Case-PRaft}

PRaft specification is divided into two modules: module \textsf{LeaderRecovery} and module \textsf{FollowerRecovery}. We conduct interaction-preserving abstraction on PRaft mainly from three perspectives, as detailed below.

%----------------------------------------
\subsection{Asynchrony Elimination} \label{ASubSec: PRaft1}

% To improve maintainability, reduce implementation complexity and support dynamic upgrade, PRaft divides control flow and IO flow by introducing a centralized coordinator which implements a set of sub protocols including leader election and follower recovery. In normal cases, servers receive and replicate commands from clients and execute those that reach consensus. Coordinator regularly checks each server's state to see whether any error occurs. When errors such as follower reboot or network partition are detected, coordinator starts corresponding handling process by sending servers commands which servers execute. Servers are passive followers and never make decisions on their own.

The centralized coordinator learns the states of servers through polling. Nodes respond to polling messages by sending replies with their system states. In implementation level specification, such polling process takes multiple steps to finish as there are several nodes in the cluster and nodes handle coordinator message asynchronously, each node responding to the polling message takes one step. When performing leader election, coordinator has to learn the checkpoint of each server. We found that the checkpoint of each server stays unchanged during polling as no valid leader exists and client commands cannot be replicated among the cluster. This suggests that the polling process is transparent to other modules. Therefore, in abstracted specification, the coordinator learns the states of all nodes synchronously in abstraction, which is safe and takes only one step. By such abstraction, traces with different permutations of polling message handling actions are all mapped to a same trace of abstracted specification.

\begin{figure}[htbp]
	\centering
	\includegraphics[width=\linewidth]{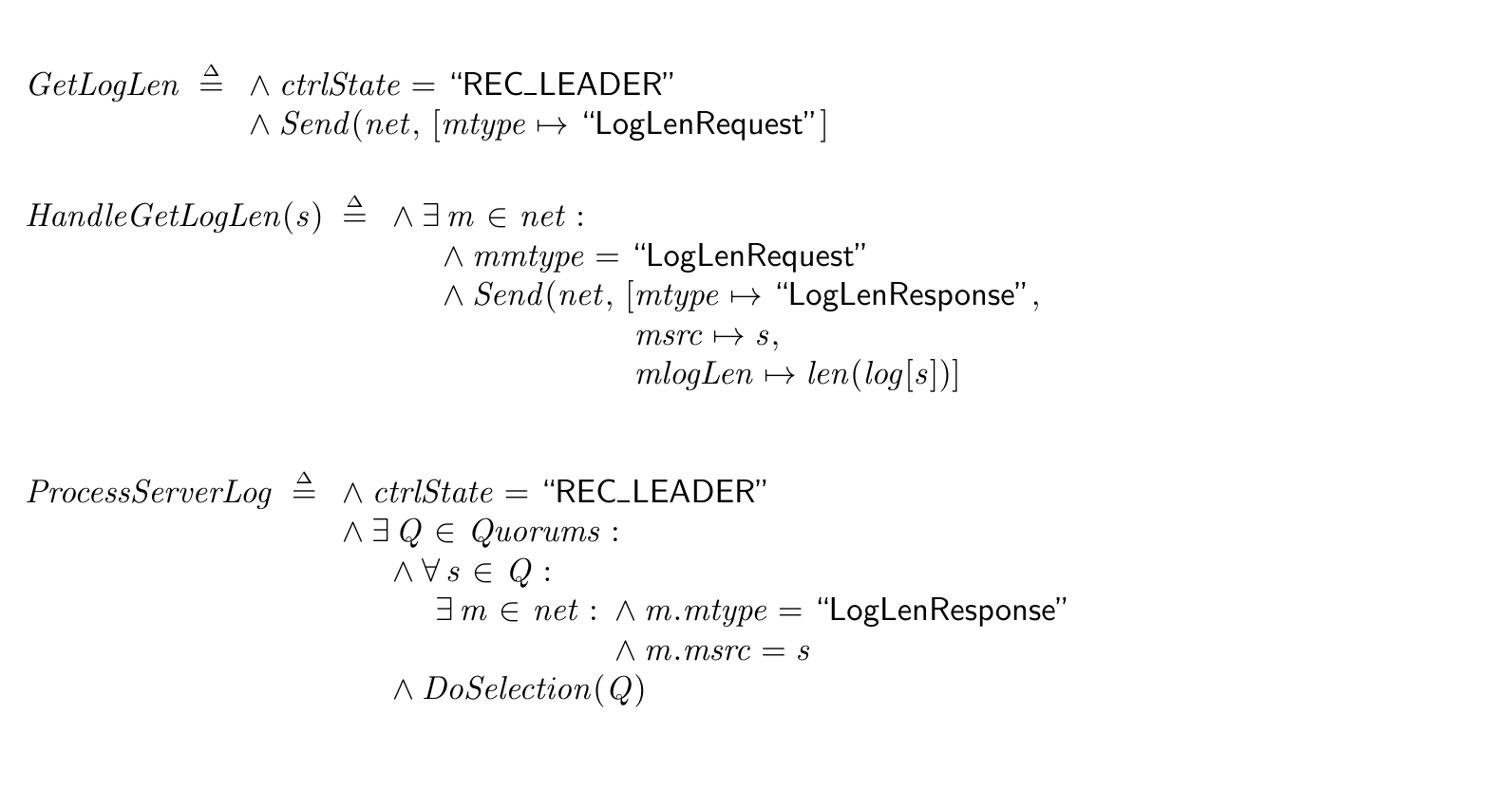}
	\caption{Specification for Selection}
	\label{F:PRaft1}
\end{figure}

Figure \ref{F:PRaft1} shows the three actions of polling process. When performing leader recovery, coordinator sends to servers requesting checkpoint of each server by action $GetLogLen$. Server handle this request by action $HandleGeyLogLen$. When coordinator receives responses from a majority of servers, it does local calculation using replies, as is specified by Action $ProcessServerLog$. We found that when state of coordinator is $REC\_LEADER$, checkpoints remain unchanged, which suggests that the order of servers handling ``LogLenRequest'' is irrelevant. 

\begin{figure}[htbp]
	\centering
	\includegraphics[width=\linewidth]{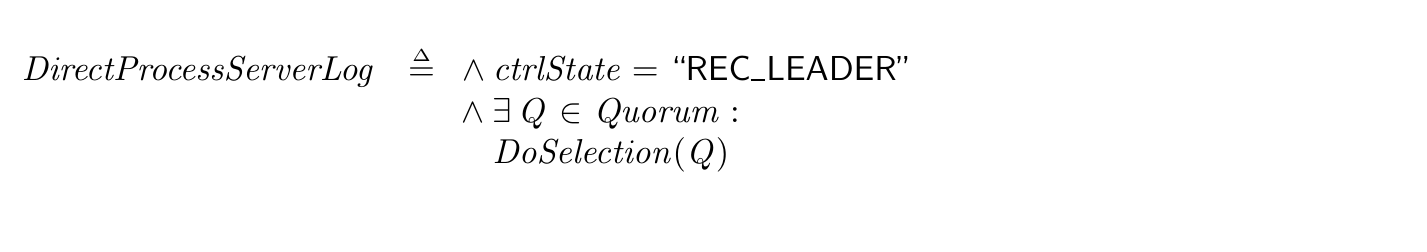}
	\caption{Abstracted Specification for Selection}
	\label{F:AbsPraft1}
\end{figure}

Figure \ref{F:AbsPraft1} shows abstracted specification. It has a single action in which coordinator directly do local calculation using globally available checkpoints of servers. This is a typical example showing how we do abstraction for one module by eliminating asynchronous behaviors that are transparent to other modules.

%----------------------------------------
\subsection{Control Flow Simplification}

%A typical control flow of PRaft follows this manner: 1. coordinator sends a command to a server. 2. server executes received command. 3. coordinator sends message to the server requesting progress. 4. server responds telling coordinator whether it has finished executing. 5. when server finishes, coordinator sends next command. Similar communications between coordinator and follower exist in both module leader recovery and module follower recovery.

%Such control procedures are suitable for system implementations but add unnecessary complexities to model checking especially the relative order between these communication steps and steps of other actions are undetermined. When performing model checking, coordinator is often redundant because a system specification stands at a global point of view and we can specify that servers make decisions on their own as if they received requests from a virtual coordinator. Thus, step 1,3 and 5 can be omitted in abstraction. We apply this abstraction in multiple control flows of PRaft.

\begin{figure}[htbp]
	\centering
	\includegraphics[width=\linewidth]{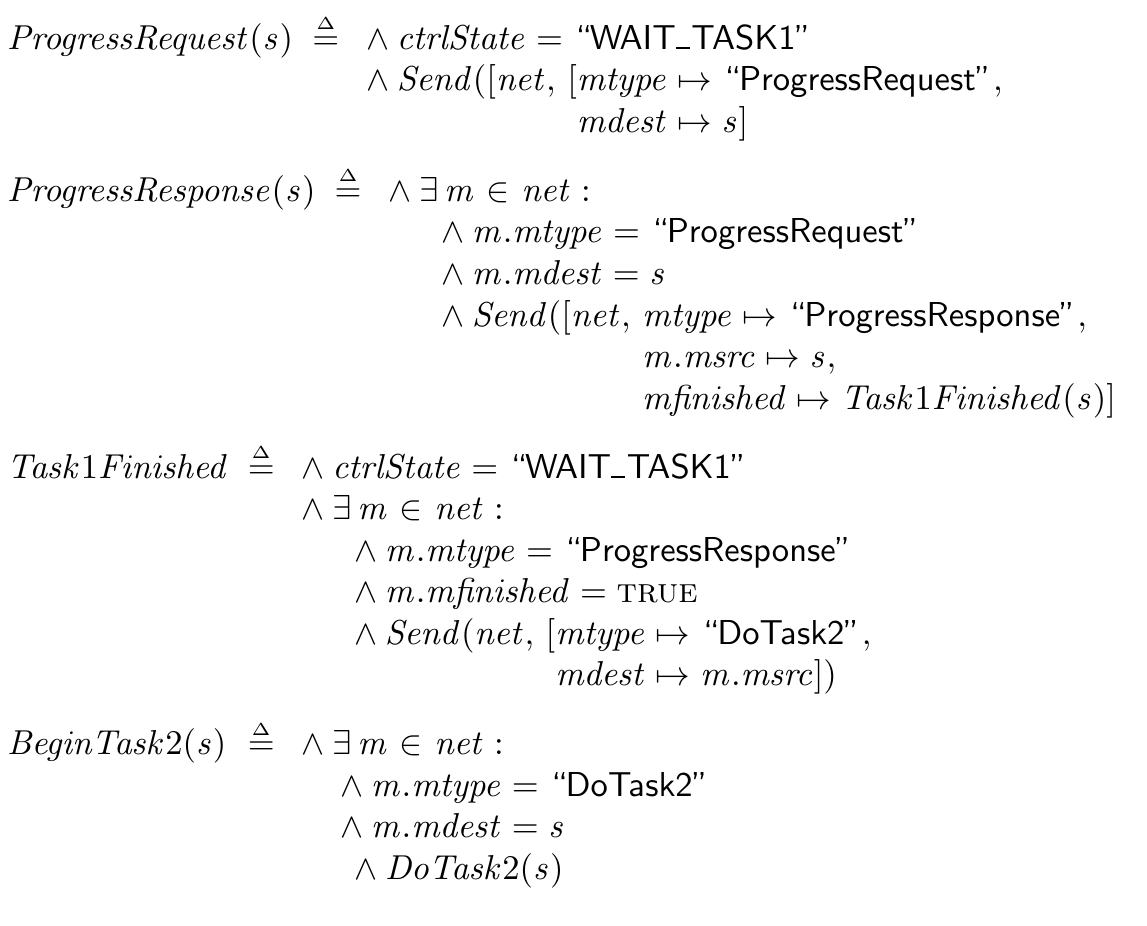}
	\caption{Specification for a Typical Control Flow}
	\label{F: Praft2}
\end{figure}

Figure \ref{F: Praft2} shows a typical control flow of PRaft. Coorinator periodically checks whether a server has finished some task it assigns as is specified by action $ProgressRequest$. When the coordinator learns from the server that it has finished, coordinator sends a message requesting the server to begin doing subsequent task. Action $Task1Finished$ specifies this process. When receiving request from coordinator, server executes command as ordered. 

\begin{figure}[htbp]
	\centering
	\includegraphics[width=\linewidth]{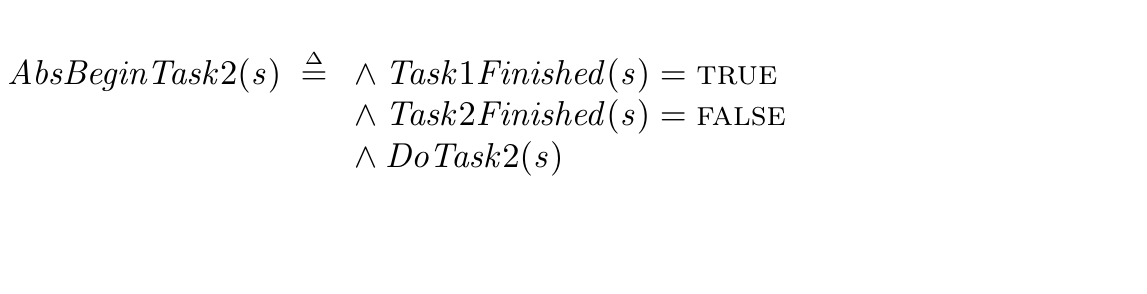}
	\caption{Abstracted Specification for Control Flow}
	\label{F: AbsRaft2}
\end{figure}

Figure \ref{F: AbsRaft2} shows the abstracted specification for this process. It omit the first three actions. When a server finishes one task, it starts doing subsequent task autonomously, as if it received an order from the coordinator. Also coordinator knows the progress of each server using global information. Therefore, the abstracted specification allows the same system behaviors as the implementation-level specification.

%----------------------------------------
\subsection{Omit Unnecessary Implementation Optimizations}

% Some optimizations in system implementations instead increase the cost of model checking. For example, in the leader recovery of PRaft, coordinator has to calculate all committed log entries from logs of a majority of servers. One way to implement is that all nodes send their logs to the coordinator, taking one round of communication. But as log is large, this may cause network congestion. To reduce network load, PRaft uses two rounds of communication. In the first round, servers simply send the length of their logs to the coordinator, who selects a majority of servers whose log is more up-to-date. In the second round, only the selected server sends their logs to the coordinator. Thus network load is reduced using one more round of communication.

\begin{figure}[htbp]
	\centering
	\includegraphics[width=\linewidth]{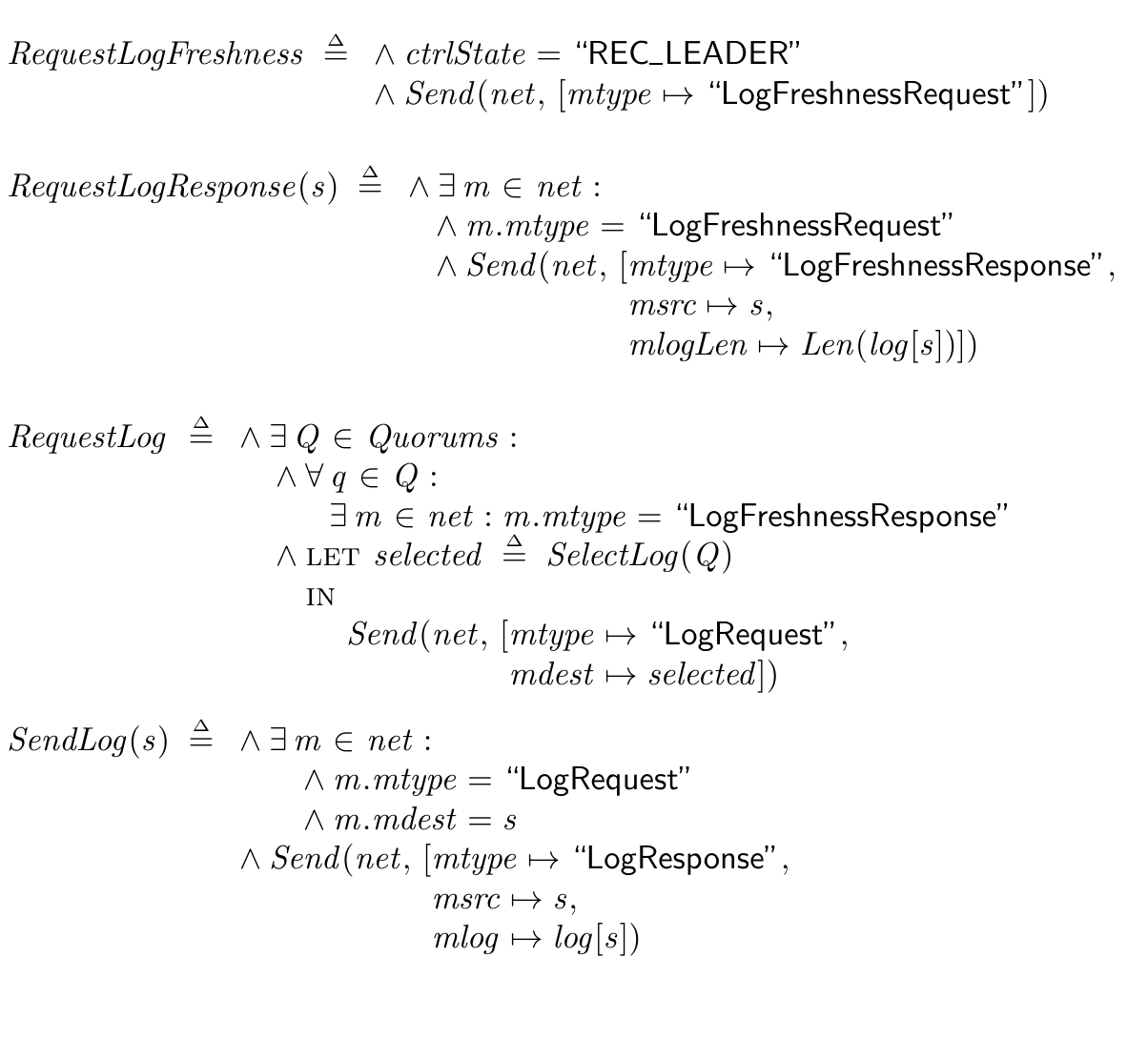}
	\caption{Specification Containing Two Rounds of Communication}
	\label{F: Praft3}
\end{figure}

Figure \ref{F: Praft3} shows the specification of coordinator collecting logs from a majority of servers using two rounds of communication. In the first round, it simply learns the length of each server's log. The first two actions specify this process. When coordinator receives replies from a majority of servers, it chooses the ones with longer logs and requests their logs, as is specified by action $RequestLog$. Note only the selected servers can receive this request. When receiving request from coordinator, server sends back its log, as is specified by action $SendLog$. Note that whether a server send coordinator the whole log or just its length makes no difference in model checking since network capacity is not considered. 

Network capacity is not modeled in specification, so it has no effect on the cost of model checking. But more rounds of communication introduce more steps in behavior traces and more possible permutations of actions, which increase the cost of model checking. Thus, we choose the solution with one round of communication in the abstracted specification.

\begin{figure}[htbp]
	\centering
	\includegraphics[width=\linewidth]{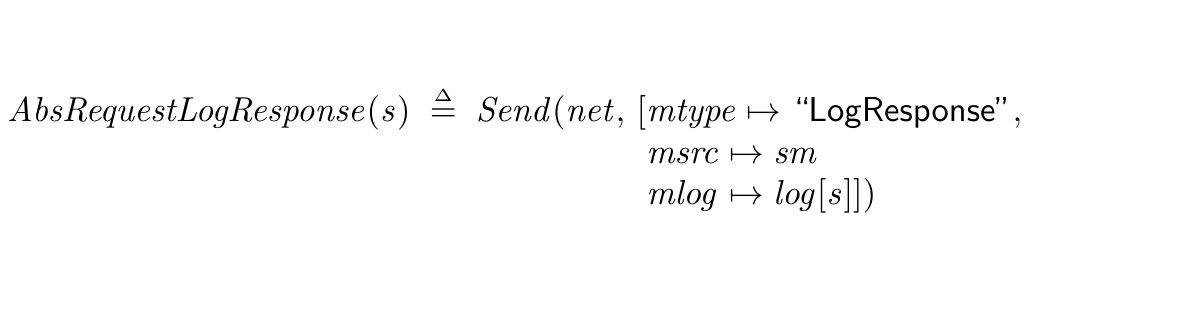}
	\caption{Abstracted Specification Containing One Round of Communication}
	\label{F: AbsRaft3}
\end{figure}

Figure \ref{F: AbsRaft3} is the abstracted specification. Each server simply sends the coordinator its whole log directly, taking only one round of communication. Note logs that coordinator receives from servers are internal variable of module leader recovery, so this difference has no influence on the other modules.

%--
%\end{CJK*}

%
% ---- Bibliography ----
%
% BibTeX users should specify bibliography style 'splncs04'.
% References will then be sorted and formatted in the correct style.
%
% \bibliographystyle{splncs04}
% \bibliography{IPA-CAV22}
%
%\begin{thebibliography}{8}
%\end{thebibliography}

\end{document}